\documentclass[preprint,aps,showpacs,floatfix]{revtex4}
\usepackage[dvips]{graphics}
\usepackage{enumerate}
\usepackage{lscape}

\begin{document}

\newcommand{\dfrac}[2]{\frac{\displaystyle #1}{\displaystyle #2}}
%%%%%%%%%%%%%%%%%%%%%%%%%%%%%%%%%%%%%%%%%%%%%%%%%%%%%%%%%%%%%%%%%
\preprint{VPI--IPPAP--04--03}
\preprint{KIAS--P04019}

\title{The NuTeV Anomaly, Lepton Universality, 
and Non-Universal Neutrino-Gauge Couplings}
\author{Will~Loinaz}\email{loinaz@alumni.princeton.edu}
\affiliation{Department of Physics, Amherst College, Amherst MA 01002}
\affiliation{Department of Physics, University of Massachusetts, Amherst MA 01003}
\author{Naotoshi~Okamura}\email{nokamura@kias.re.kr}
\affiliation{Korea Institute for Advanced Study, 
207-43 Cheongnyangni 2-dong, Dongdaemun-gu, 
Seoul 130-722, Korea}
\author{Saifuddin~Rayyan}\email{srayyan@vt.edu}
\author{Tatsu~Takeuchi}\email{takeuchi@vt.edu}
\affiliation{Institute for Particle Physics and Astrophysics,
Physics Department, Virginia Tech, Blacksburg VA 24061}
\author{L.~C.~R.~Wijewardhana}\email{rohana@physics.uc.edu}
\affiliation{Department of Physics, University of Cincinnati, Cincinnati OH 45221-0011}

\date{March 31, 2004, Revised June 29, 2004}

\begin{abstract}
\noindent

In previous studies we found that models with flavor-universal suppression of the 
neutrino-gauge couplings are compatible with NuTeV and $Z$-pole data.  
In this paper we expand our analysis to obtain constraints on 
flavor-dependent coupling suppression by including lepton universality data
from $W$, $\tau$, $\pi$ and $K$ decays in fits to model parameters.
We find that the data are consistent with a variety of patterns of coupling suppression.
In particular, in scenarios in which the suppression arises from the mixing of 
light neutrinos with heavy gauge singlet states (neutrissimos), we find patterns 
of flavor-dependent coupling suppression which are also consistent with 
constraints from $\mu \rightarrow e \gamma$.
\end{abstract}


\maketitle
%%%%%%%%%%%%%%%%%%%%%%%%%%%%%%%%%%%%%%%%%%%%%%%%%%%%%%%%%%%%%%%%%
\section{Introduction}

Recent analysis of $\nu_\mu$ ($\bar{\nu}_\mu$) scattering data from
the NuTeV experiment at Fermilab \cite{Zeller:2001hh} 
indicates a value of the effective neutrino-quark coupling parameter $g_L^2$ 
which deviates by $3 \sigma$ from the Standard Model prediction (based on a global fit using non-NuTeV data).  
The significance of the NuTeV result remains controversial \cite{Davidson:2001ji}, 
and a critical examination of the initial analysis is ongoing.  
Several groups are evaluating potential theoretical 
uncertainties arising from purely Standard Model physics which might be comparable to or 
larger than the quoted experimental uncertainty of the NuTeV result.  
Candidate sources of large theoretical uncertainty include 
next-to-leading-order (NLO) QCD corrections \cite{Dobrescu:2003}, 
NLO electroweak corrections \cite{Diener:2003}, and parton distribution functions 
(especially as involves assumptions about sea-quark asymmetries) \cite{Gambino:2003}.  
The effect of the first has been estimated to be comparable in size to the NuTeV experimental 
uncertainty, while the latter two might give rise to effects comparable in size to the 
full NuTeV discrepancy with the Standard Model.  Elucidation of the actual impact of 
these effects on the NuTeV result awaits a reanalysis of the NuTeV data.  
However, it remains a distinct possibility that the discrepancy with the 
Standard Model prediction is genuine and that its resolution lies in physics 
beyond the Standard Model \cite{Chanowitz:2002}.  
It is this possibility that we investigate here.

In a previous paper \cite{LOTW1}, we demonstrated that the
$Z$-pole data from $e^+e^-$ colliders \cite{LEP/SLD:2003}
and the $\nu_\mu$ ($\bar{\nu}_\mu$) scattering data from NuTeV \cite{Zeller:2001hh}
are compatible if (1) the Higgs boson is heavy and (2)
the $Z\nu_\ell\nu_\ell$  and $W\ell\nu_\ell$  ($\ell=e,\mu,\tau$) 
couplings are suppressed by a factors 
 $(1-\varepsilon_\ell)$  and $(1-\varepsilon_\ell/2)$, respectively. 
We also showed that such suppressions could arise from
neutrinos mixing with heavy gauge singlet (neutrissimo) states \cite{numix,Shrock:1977,Glashow:2003wt,LORTW2}.

In Ref.~\cite{LOTW1}, it was assumed that the suppression parameters
were flavor universal: 
$\varepsilon_e = \varepsilon_\mu = \varepsilon_\tau \equiv \varepsilon$.
The value of $\varepsilon$ required to fit the data was 
\begin{equation}
\varepsilon = 0.0030 \pm  0.0010 \;.
\label{epslimit}
\end{equation}
However, in seesaw models \cite{seesaw} of neutrino masses and mixings
such a large universal $\varepsilon$ implies a
prohibitively large rate of $\mu\rightarrow e\gamma$ \cite{Shrock:1977,Glashow:2003wt,LORTW2}.
To bring the models into agreement with experiment the assumption of universality must be relaxed:
either $\epsilon_e$ or $\epsilon_\mu$, but not both, 
must be strongly suppressed \footnote{Note that this restriction 
applies only to seesaw models with equal numbers of sterile and active 
neutrinos.}.  
Further, in most models flavor-universal suppressions
require considerable fine tuning.  It is thus natural to ask what patterns of flavor non-universal
suppressions are consistent with the data.  
If the suppression parameters can be flavor-dependent, one must also ask
whether the preferred values of the $\varepsilon_\ell$ are all positive,  
\textit{i.e.} are all the neutrino-gauge couplings suppressed?
Negative $\varepsilon_\ell$ indicates an enhancement of the
$W\ell\nu_\ell$ and $Z\nu_\ell\nu_\ell$ couplings which cannot be arranged
via neutrino mixing. 

In addition to the $Z$-pole and NuTeV data, there is a
wealth of experimental data bounding lepton universality violation in the charged
channel from $W$, $\pi$, $K$, and $\tau$--decays \cite{Shrock}. 
In the following, we analyze the constraints that 
$Z$-pole and NuTeV data and the lepton universality bounds
impose on neutrino-mixing models by fitting the data with
flavor-dependent suppression parameters $\varepsilon_\ell$ 
($\ell = e,\mu,\tau$) along with the $S$, $T$, and $U$ oblique
correction parameters \cite{Peskin:1990zt}.
We perform fits in which all six parameters float independently,
and we also fit to models in which one or more of the  $\varepsilon_\ell$ are 
assumed to be strongly suppressed.  As in the flavor-universal case, the data
require a negative $T$ parameter and a positive $U$ parameter.
However, we find that the data are consistent with a variety of
patterns of suppression parameters, including patterns compatible with
$\mu \rightarrow e \gamma$ data.

\section{Constraints on Lepton Universality}

Here, we survey current experimental constraints on lepton 
universality. For a comprehensive review on the subject, see
Ref.~\cite{LepUnivRev}.
We parametrize the couplings of the $W^\pm$'s with the leptons as
\begin{equation}
\mathcal{L} = \sum_{\ell=e,\mu,\tau}
\frac{g_\ell}{\sqrt{2}}\, W^+_\mu \,\bar{\nu}_\ell\, 
\gamma^\mu\left(\frac{1-\gamma_5}{2}\right)\ell^- + \mathrm{h.c.}\;.
\end{equation}
The Standard Model assumes $g_e=g_\mu=g_\tau=g$. 
Although experimental limits on the ratios $g_\mu/g_e$,
$g_\tau/g_\mu$, and $g_\tau/g_e$ have been calculated and tabulated
as recently as fall 2002 by Pich \cite{Pich:2002bc}, we repeat 
the exercise here to incorporate more recent data and to obtain
the correlations among the limits necessary for our analysis.

%%%%%%%%%%%%%%%%%%%%%%%%%%%%%%%%%%%%%%%%%%%%%

\subsection{$W$-decay}

\noindent
The decay width of the $W$ at tree level is
\begin{equation}
\Gamma(W\rightarrow \ell\,\bar{\nu}_\ell)
= \frac{g_\ell^2 M_W}{48\pi}
  \left( 1-\frac{m_\ell^2}{M_W^2}\right)^2
  \left( 1+\frac{m_\ell^2}{2M_W^2}\right)\;.
\label{GammaW}
\end{equation}
The branching fractions of the $W$ into the three lepton generations
have been measured at LEP~II to be (Ref.~\cite{LEP/SLD:2003}, page 74)
\begin{eqnarray}
B(W\rightarrow e\bar{\nu}_e) & = & 10.59\pm 0.17\%\;,\cr
B(W\rightarrow \mu\bar{\nu}_\mu) & = & 10.55\pm 0.16\%\;,\cr
B(W\rightarrow \tau\bar{\nu}_\tau) & = & 11.20\pm 0.22\%\;,
\end{eqnarray}
with correlations shown in Table~\ref{WdecayCorr}.
From this data, we find (Ref.~\cite{LEP/SLD:2003}, page 73)
\begin{eqnarray}
B(W\rightarrow \mu\bar{\nu}_\mu)/
B(W\rightarrow e\bar{\nu}_e) & = & 0.997 \pm 0.021\;, \cr
B(W\rightarrow \tau\bar{\nu}_\tau)/
B(W\rightarrow e\bar{\nu}_e) & = & 1.058 \pm 0.029\;, 
\label{WdecayData}
\end{eqnarray}
with a correlation of $+0.44$ between the two ratios.
Using Eq.~(\ref{GammaW}), this translates into
\begin{eqnarray}
(g_\mu/g_e)_W  & = & 0.999\pm 0.011\;,\cr
(g_\tau/g_e)_W & = & 1.029\pm 0.014\;,
\end{eqnarray}
with a correlation of $+0.44$.
The central values have shifted slightly from Ref.~\cite{Pich:2002bc}
due to the update of the $W$ branching fractions from LEP~II.

\begin{table}[ht]
\begin{tabular}{l|ccc}
& $B(W\rightarrow e\bar{\nu}_e)$ 
& $B(W\rightarrow \mu\bar{\nu}_\mu)$ 
& $B(W\rightarrow \tau\bar{\nu}_\tau)$ \\
\hline
$B(W\rightarrow e\bar{\nu}_e)$       & $1.000$ & $0.092$ & $-0.196$ \\
$B(W\rightarrow \mu\bar{\nu}_\mu)$   &         & $1.000$ & $-0.148$ \\
$B(W\rightarrow \tau\bar{\nu}_\tau)$ &         &         & $\phantom{-}1.000$ \\
\hline
\end{tabular}
\caption{Correlations among the $W$ branching fractions 
measured at LEP~II (Ref.~\cite{LEP/SLD:2003}, page 182).}
\label{WdecayCorr}
\end{table}
%

%%%%%%%%%%%%%%%%%%%%%%%%%%%%%%%%%%%%%%%%%%%%%

\subsection{$\tau$ and $\mu$ decay}

\noindent
The decay widths of the $\tau$ and $\mu$ into 
lighter leptons, including radiative corrections \cite{Kinoshita:1958ru,Marciano:vm}, 
are:
\begin{eqnarray}
\Gamma(\tau\rightarrow \mu\, \bar{\nu}_\mu\, \nu_\tau\,(\gamma) ) 
& = & \frac{g_\tau^2 g_\mu^2}{64 M_W^4}\;
      \frac{ m_\tau^5 }{ 96\pi^3}\;
       f\!\left(\frac{m_\mu^2}{m_\tau^2}\right)
      \delta^\tau_W\, \delta^\tau_\gamma\;, \cr
\Gamma(\tau\rightarrow e\, \bar{\nu}_e\, \nu_\tau\,(\gamma) ) 
& = & \frac{g_\tau^2 g_e^2}{64 M_W^4}\;
      \frac{ m_\tau^5 }{ 96\pi^3}\;
       f\!\left(\frac{m_e^2}{m_\tau^2}\right)
      \delta^\tau_W\, \delta^\tau_\gamma\;, \cr
\Gamma(\mu\rightarrow e\, \bar{\nu}_e\, \nu_\mu\,(\gamma) ) 
& = & \frac{g_\mu^2 g_e^2}{64 M_W^4}\;
      \frac{ m_\mu^5 }{ 96\pi^3}\;
       f\!\left(\frac{m_e^2}{m_\mu^2}\right)
      \delta^\mu_W\, \delta^\mu_\gamma
\;=\; \frac{1}{\tau_\mu}\;, 
\end{eqnarray}
in which $f(x)$ is the phase space factor
\begin{equation}
f(x) = 1 - 8x + 8x^3 - x^4 - 12 x^2\ln x\;,
\end{equation}
$\delta^\ell_W$ is the $W$ propagator correction
\begin{equation}
\delta^\ell_W = \left( 1 +\frac{3}{5}\frac{m_\ell^2}{M_W^2}
                \right),
\end{equation}
$\delta^\ell_\gamma$ is the radiative correction from photons
\begin{equation}
\delta^\ell_\gamma = 
1 + \frac{\alpha(m_\ell)}{2\pi}\left( \frac{25}{4} - \pi^2 \right)\;,
\end{equation}
and the values of the running QED coupling constant at
relevant energies are \cite{Marciano:vm}
\begin{eqnarray}
\alpha^{-1}(m_\mu) & = & \alpha^{-1} 
-\frac{2}{3\pi}\ln\frac{m_\mu}{m_e} 
+\frac{1}{6\pi}
\approx 136.0\;, \cr
\alpha^{-1}(m_\tau) & \approx & 133.3 \;.
\end{eqnarray}
The numerical values of these corrections are shown in Table~\ref{taumucorrections}.
\begin{table}
\begin{tabular}{l||c|c|c}
& phase space & $W$ propagator & photon \\
\hline
$\Gamma(\tau\rightarrow \mu\,\bar{\nu}_\mu\, \nu_\tau)$ &
$\;f(m_\mu^2/m_\tau^2) = 0.9726\;$ & 
$\;\delta^\tau_W = 1.0003\;$ & 
$\;\delta^\tau_\gamma = 0.9957\;$ \\
\cline{1-2}
$\Gamma(\tau\rightarrow e\,\bar{\nu}_e\, \nu_\tau)$ &
$\;f(m_e^2/m_\tau^2) = 1.0000\;$ & & \\
\hline
$\Gamma(\mu\rightarrow e\,\bar{\nu}_e\, \nu_\mu)$ &
$\;f(m_e^2/m_\mu^2) = 0.9998\;$ & 
$\;\delta^\mu_W = 1.0000\;$ & 
$\delta^\mu_\gamma = 0.9958$ \\
\hline
\end{tabular}
\caption{The corrections to the leptonic decay widths of the $\tau$ and $\mu$.}
\label{taumucorrections}
\end{table}
The ratios of the coupling constants can be extracted using the relations
\begin{eqnarray}
\frac{\Gamma(\tau\rightarrow\mu\,\bar{\nu}_\mu\,\nu_\tau\,(\gamma))}
     {\Gamma(\tau\rightarrow e\,\bar{\nu}_e\,\nu_\tau\,(\gamma))}
& = &
\frac{B(\tau\rightarrow\mu\,\bar{\nu}_\mu\,\nu_\tau\,(\gamma))}
     {B(\tau\rightarrow e\,\bar{\nu}_e\,\nu_\tau\,(\gamma))}
\;=\; \frac{g_\mu^2}{g_e^2}\;
      \frac{f(m_\mu^2/m_\tau^2)}
           {f(m_e^2  /m_\tau^2)}\;,\cr
\frac{\Gamma(\tau\rightarrow\mu\,\bar{\nu}_\mu\,\nu_\tau\,(\gamma))}
     {\Gamma(\mu\rightarrow e\,\bar{\nu}_e\,\nu_\mu\,(\gamma))}
& = &
\frac{\tau_\mu}{\tau_\tau}\;
B(\tau\rightarrow\mu\,\bar{\nu}_\mu\,\nu_\tau\,(\gamma))
\;=\; \frac{g_\tau^2}{g_e^2}\;
      \frac{m_\tau^5}{m_\mu^5}\;
      \frac{f(m_\mu^2/m_\tau^2)}
           {f(m_e^2  /m_\mu^2)}\;
      \frac{\delta^\tau_W}{\delta^\mu_W}\;
      \frac{\delta^\tau_\gamma}{\delta^\mu_\gamma}\;.
\end{eqnarray}
The latest world averages for the quantities 
appearing in these equations are listed in Table~\ref{MassLifeBranch}, 
which yield
\begin{eqnarray}
(g_\mu/g_e)_{\tau\phantom{\mu}} & = & 0.9999\pm 0.0021 \;,\cr
(g_\tau/g_e)_{\tau\mu} & = & 1.0004\pm 0.0022 \;,
\label{taudecay}
\end{eqnarray}
with a correlation of $0.51$ due to the
inputs $B(\tau\rightarrow\mu\,\bar{\nu}_\mu\,\nu_\tau\,(\gamma))$
and $m_\tau$ common to both ratios.

\begin{table}[t]
\begin{tabular}{l|l|c}
Observable & \ World Average & \ Reference \\
\hline
$m_e$ (MeV)     & $\;0.510998918\pm 0.000000044$         & \cite{PDG2004} \\
\hline
$m_\mu$ (MeV)   & $\;105.6583692\pm 0.0000094$           & \cite{PDG2004} \\
$\tau_\mu$ (s)  & $\;(2.19703\pm 0.00004)\times 10^{-6}\;$ & \cite{PDG2004} \\
\hline
$m_\tau$ (MeV)  & $\;1776.99{}^{+0.29}_{-0.26}$          & \cite{PDG2004} \\
$\tau_\tau$ (s) & $\;(290.6\pm 0.9)\times 10^{-15}$      & \cite{Gan:2003fj} Figure~1  \\
$B(\tau\rightarrow e\,\bar{\nu}_e\,\nu_\tau)\;$
& $\;17.823\pm 0.051\,\%$  &  \cite{Davier:2002mn} Figure~8 \\
$B(\tau\rightarrow\mu\,\bar{\nu}_\mu\,\nu_\tau)\;$
& $\;17.331\pm 0.054\,\%$  &  \cite{Davier:2002mn} Figure~9 \\
$B(\tau\rightarrow \pi\,\nu_\tau)\;$
& $\;10.975\pm 0.065\,\%$  
& \cite{PDG2004}, \cite{Gan:2003fj} Figure~3, \cite{Davier:2002mn} Table~3 \\
$B(\tau\rightarrow K\,\nu_\tau)\;$
& $\;0.686\pm 0.023\,\%\;$ &  \cite{PDG2004} \\
\hline
$m_\pi$ (MeV)   & $\;139.57018\pm 0.00035$               & \cite{PDG2004} \\
$\tau_\pi$ (s)  & $\;(2.6033\pm 0.0005)\times 10^{-8}$   & \cite{PDG2004} \\
$B(\pi\rightarrow \mu\,\bar{\nu}_\mu)\;$
& $\;99.98770\pm 0.00004\,\%$ & \cite{PDG2004} \\
$B(\pi\rightarrow e\,\bar{\nu}_e)\;$
& $\;(1.230\pm 0.004)\times 10^{-4}\;$ & \cite{PDG2004} \\
%& $\;(1.2310\pm 0.0037)\times 10^{-4}\;$ & \cite{Britton:1992pg}, \cite{Czapek:kc} \\
\hline
$m_K$ (MeV)     & $\;493.677\pm 0.016\;$ & \cite{PDG2004} \\
$\tau_K$ (s)    & $\;(1.2384\pm 0.0024)\times 10^{-8}\;$ & \cite{PDG2004} \\
$B(K\rightarrow \mu\,\bar{\nu}_\mu)\;$
& $\;63.43\pm 0.17\,\%$ & \cite{PDG2004} \\
%$B(K\rightarrow e\,\bar{\nu}_e)\;$
%& $\;(1.55\pm 0.07)\times 10^{-5}\;$ & \cite{Hagiwara:fs} \\
\hline
\end{tabular}
\caption{The world averages of masses, life times, and branching fractions
used in this analysis.  The branching fractions subsume the decays with $\gamma$'s.}
\label{MassLifeBranch}
\end{table}
%

%%%%%%%%%%%%%%%%%%%%%%%%%%%%%%%%%%%%%%%%%%%%%

\begin{table}[t]
\begin{tabular}{l|l}
Processes & \ Constraint \\
\hline
\hline
$W\rightarrow e\,\bar{\nu}_e\;$  
& $\;(g_\mu/g_e)_W  = 0.999\pm 0.011\;$ \\
$W\rightarrow \mu\,\bar{\nu}_\mu\;$  
& $\;(g_\tau/g_e)_W = 1.029\pm 0.014\;$ \\
$W\rightarrow \tau\,\bar{\nu}_\tau\;$  & \\
\hline
$\mu\rightarrow e\,\bar{\nu}_e\,\nu_\mu\;$ 
& $\;(g_\mu/g_e)_{\tau\phantom{\mu}} = 0.9999\pm 0.0021\;$ \\
$\tau\rightarrow e\,\bar{\nu}_e\,\nu_\tau\;$ 
& $\;(g_\tau/g_e)_{\tau\mu}  = 1.0004\pm 0.0022\;$ \\
$\tau\rightarrow \mu\,\bar{\nu}_\mu\,\nu_\tau\;$ & \\
\hline
$\pi\rightarrow \mu\,\bar{\nu}_\mu\;$ 
& $\;(g_\mu/g_e)_{\pi\phantom{\tau}} = 1.0021\pm 0.0016\;$ \\
$\pi\rightarrow e\,\bar{\nu}_e\;$ 
& $\;(g_\tau/g_\mu)_{\pi\tau} = 1.0030\pm 0.0034\;$ \\
$\tau\rightarrow \pi\,\nu_\tau\;$ & \\
\hline
$K\rightarrow \mu\,\bar{\nu}_\mu\;$ 
& $\;(g_\tau/g_\mu)_{K\tau} = 0.979\pm 0.017\;$ \\
$\tau\rightarrow K\,\nu_\tau\;$ & \\
\hline
\end{tabular}
\caption{Limits on lepton universality from various processes.}
\label{LeptUnivConst}
\end{table}

%%%%%%%%%%%%%%%%%%%%%%%%%%%%%%%%%%%%%%%%%%%%%

\subsection{Pion and $\tau$ decay}

\noindent
At tree level,
the widths of charged $\pi$-decay into leptons are
\begin{eqnarray}
\Gamma(\pi\rightarrow e\,\bar{\nu}_e)
& = & \frac{g_e^2 g_{ud}^2}{256\pi}\,\frac{f_\pi^2}{M_W^4}\,
      m_e^2 m_\pi\left(1-\frac{m_e^2}{m_\pi^2}\right)^2\;, \cr
\Gamma(\pi\rightarrow \mu\,\bar{\nu}_\mu)
& = & \frac{g_\mu^2 g_{ud}^2}{256\pi}\,\frac{f_\pi^2}{M_W^4}\,
      m_\mu^2 m_\pi\left(1-\frac{m_\mu^2}{m_\pi^2}\right)^2\,\;,
\end{eqnarray}
while that of $\tau$-decay into $\pi\,\nu_\tau$ is
\begin{equation}
\Gamma(\tau\rightarrow \pi\,\nu_\tau)
= \frac{g_\tau^2 g_{ud}^2}{512\pi}\,\frac{f_\pi^2}{M_W^4}\,
  m_\tau^3\left(1-\frac{m_\pi^2}{m_\tau^2}\right)^2\;,
\end{equation}
where $g_{ud} = g|V_{ud}|$, and the pion decay constant $f_\pi$ is normalized
as (Ref.~\cite{Hagiwara:fs}, page 439)
\begin{equation}
\langle 0 |\,\bar{u}\gamma_\mu \gamma_5 d \,(0)\,| \pi^-(\mathbf{q}) \rangle = i\,q_\mu f_\pi\;.
\end{equation}
Taking ratios, we find
\begin{eqnarray}
R^0_{e/\mu} \;\equiv\;
\frac{\Gamma(\pi\rightarrow e \,\bar{\nu}_e  )}
     {\Gamma(\pi\rightarrow\mu\,\bar{\nu}_\mu)}
& = &
\frac{B(\pi\rightarrow e \,\bar{\nu}_e  )}
     {B(\pi\rightarrow\mu\,\bar{\nu}_\mu)}
\;=\;
\frac{g_e^2}{g_\mu^2}\,
\frac{m_e^2}{m_\mu^2}\,
\frac{(1-m_e^2  /m_\pi^2)^2}
     {(1-m_\mu^2/m_\pi^2)^2} \;, \cr
R^0_{\tau/\pi} \;\equiv\;
\frac{\Gamma(\tau\rightarrow\pi\,\nu_\tau     )}
     {\Gamma(\pi \rightarrow\mu\,\bar{\nu}_\mu)}
& = &
\frac{\tau_\pi}{\tau_\tau}\,
\frac{B(\tau\rightarrow\pi \,\nu_\tau    )}
     {B(\pi \rightarrow\mu\,\bar{\nu}_\mu)}
\;=\;
\frac{g_\tau^2}{g_\mu^2}\,
\frac{m_\tau^3}{2 m_\mu^2 m_\pi}\;
\frac{(1-m_\pi^2/m_\tau^2)^2}
     {(1-m_\mu^2 /m_\pi^2)^2} \;.
\end{eqnarray}
Radiative corrections to these relations have been calculated
in Ref.~\cite{Decker:1994ea} and modify them to
\begin{eqnarray}
R_{e/\mu} & = & 
\frac{B(\pi\rightarrow e \,\bar{\nu}_e  \,(\gamma))}
     {B(\pi\rightarrow\mu\,\bar{\nu}_\mu\,(\gamma))}
\;=\;
R^0_{e/\mu}
\left(1+\delta R_{e/\mu}\right) \;, \cr
R_{\tau/\pi} & = &
\frac{\tau_\pi}{\tau_\tau}\,
\frac{B(\tau\rightarrow\pi \,\nu_\tau     \,(\gamma))}
     {B(\pi \rightarrow\mu\,\bar{\nu}_\mu \,(\gamma))}
\;=\;
R^0_{\tau/\pi}
\left(1+\delta R_{\tau/\pi}\right) \;,
\end{eqnarray}
with
\begin{equation}
\delta R_{e/\mu} = -0.0374\pm 0.0001 \;,\qquad
\delta R_{\tau/\pi} = +0.0016^{+0.0009}_{-0.0014}\;.
\end{equation}
The uncertainty in these corrections is due to
the uncertainty from strong interaction effects.
With these relations and the experimental data listed 
in Table~\ref{MassLifeBranch}, we obtain
\begin{eqnarray}
(g_\mu/g_e)_{\pi\phantom{\tau}} & = & 1.0021\pm 0.0016 \;,\cr
(g_\tau/g_\mu)_{\pi\tau} & = & 1.0030\pm 0.0034 \;.
\label{pidecay}
\end{eqnarray}
The correlation between the two is virtually zero due to the
accuracy of the common inputs $m_\mu$, $m_\pi$, and
$B(\pi\rightarrow\mu\,\bar{\nu}_\mu)$.
There is a correlation of $+0.33$ between $(g_\tau/g_\mu)_{\pi\tau}$ and
$(g_\tau/g_e)_{\tau\mu}$ of Eq.~(\ref{taudecay}) arising from the common inputs
$\tau_\tau$ and $m_\tau$.
A few comments are in order:
\begin{itemize}
\item
Our limit on $(g_\mu/g_e)_{\pi}$ differs from that of Pich and Silva \cite{Pich:1995vj} 
who use for the value of $B(\pi\rightarrow e\,\bar{\nu}_e (\gamma))$
the weighted average of the results from TRIUMF \cite{Britton:1992pg}
and PSI \cite{Czapek:kc}, $(1.2310\pm 0.0037)\times 10^{-4}$.
If we use this value
instead of the average from the Review of Particle Properties
\cite{PDG2004} listed in Table~\ref{MassLifeBranch}, 
we obtain $(g_\mu/g_e)_{\pi} = 1.0017\pm 0.0015$
in agreement with Ref.~\cite{Pich:1995vj}.

\item
The experimental value of $B(\tau\rightarrow \pi\nu_\tau)$ listed in 
Table~\ref{MassLifeBranch} is the average of CLEO and the four LEP experiments.
CLEO \cite{cleo}, OPAL \cite{opal}, DELPHI \cite{delphi}, and L3 \cite{l3} 
report the semi-exclusive branching fraction
$B(\tau\rightarrow h\nu_\tau)$, where $h=\pi$ or $K$, as
\begin{equation}
\begin{array}{llclclcll}
\mathrm{CLEO:} & B(\tau\rightarrow h\nu_\tau) & = & 11.52  &\pm & 0.05  &\pm & 0.12 &\% \cr
\mathrm{OPAL:} &                              & = & 11.98  &\pm & 0.13  &\pm & 0.16 &\% \cr
\mathrm{DELPHI:} \qquad &                     & = & 11.601 &\pm & 0.120 &\pm & 0.116 &\% \cr
\mathrm{L3:}   &                              & = & 12.09  &\pm & 0.12  &\pm & 0.10  &\% 
\end{array}
\end{equation}
The CLEO and OPAL values are published and 
used in the average of the Review of Particle Properties \cite{PDG2004}.
Adding the statistical and systematic errors in quadrature and taking the weighted average
of these four numbers, we obtain
\begin{equation}
\mbox{average without ALEPH:}\qquad B(\tau\rightarrow h\nu_\tau) = 11.752\pm 0.079\;\%\;.
\end{equation}
As noted by Gan \cite{Gan:2003fj}, the agreement among these four measurements is poor ($\chi^2/\mathrm{d.o.f.}=9.9/3$, where d.o.f. is degrees of freedom). 
This stands in stark constrast to the situation in fall 2002 
when the agreement among CLEO and the four LEP experiments was much better ($\chi^2/\mathrm{d.o.f.}=5.09/4$) \cite{Matorras:2002yx}.
The source of the difference is the new L3 value \cite{l3} which has a much higher central value and
smaller error bar than before \cite{Acciarri:1994vr}.
Subtracting the world average $B(\tau\rightarrow K\nu_\tau) = 0.686\pm 0.023\%$ \cite{PDG2004}, we obtain
\begin{equation}
\mbox{average without ALEPH:}\qquad
B(\tau\rightarrow \pi\nu_\tau) = 11.066\pm 0.082\;\%\;.
\label{CODL}
\end{equation}
ALEPH (\cite{Davier:2002mn}, Table~3) reports the value of the exclusive branching fraction
$B(\tau\rightarrow \pi\nu_\tau)$ as
\begin{equation}
\mathrm{ALEPH:}\qquad
B(\tau\rightarrow \pi\nu_\tau) = 10.828\pm 0.070\pm 0.078\;\%\;.
\end{equation}
which does not agree particularly well with Eq.~(\ref{CODL}) either.
Although this ALEPH value is excluded from the world average by Gan 
in Ref.~\cite{Gan:2003fj} as preliminary, 
we include it in our analysis since it was included in the previous
analysis by Pich \cite{Pich:2002bc} (with the caveat that it is subject to change).
The weighted average with Eq.~(\ref{CODL}) is
\begin{equation}
\mbox{world average:}\qquad
B(\tau\rightarrow \pi\nu_\tau) = 10.975\pm 0.065\;\%\;,
\end{equation}
which is the value used to obtain Eq.~(\ref{pidecay}). The associated 
$\chi^2/\mathrm{d.o.f.}$ is 13.1/4, so is 
unimproved with the inclusion of the ALEPH result.   
If we exclude the ALEPH value and use Eq.~(\ref{CODL}) instead, we obtain
$(g_\tau/g_\mu)_{\pi\tau} = 1.0072\pm 0.0041$.

The current state of agreement among the data determining $B(\tau\rightarrow \pi\nu_\tau)$ is clearly unsatisfactory.  Additional data, perhaps from new experiments at CLEO \cite{Weinstein:2002nh}, are needed to provide a definitive value.

\end{itemize}
%%%%%%%%%%%%%%%%%%%%%%%%%%%%%%%%%%%%%%%%%%%%%

\begin{table}[t]
\begin{tabular}{l||cc|ccccc}
& $(g_\mu/g_e)_W$
& $(g_\tau/g_e)_W$ 
& $(g_\mu/g_e)_\tau$ 
& $(g_\tau/g_e)_{\tau\mu}$
& $(g_\mu/g_e)_\pi$
& $(g_\tau/g_\mu)_{\pi\tau}$ 
& $(g_\tau/g_\mu)_{K\tau}$ \\
\hline\hline
$(g_\mu/g_e)_W$  & 1.00 & 0.44 &  &  &  &  &  \\
$(g_\tau/g_e)_W$ &      & 1.00 &  &  &  &  &  \\
\hline
$(g_\mu/g_e)_\tau$ &    &      & 1.00 & 0.51 & 0.00 & 0.00 & 0.00 \\
$(g_\tau/g_e)_{\tau\mu}$ &  &  &      & 1.00 & 0.00 & 0.33 & 0.07 \\
$(g_\mu/g_e)_\pi$        &  &  &      &      & 1.00 & 0.00 & 0.00 \\
$(g_\tau/g_\mu)_{\pi\tau}$ & & &      &      &      & 1.00 & 0.04 \\
$(g_\tau/g_\mu)_{K\tau}$   & & &      &      &      &      & 1.00 \\
\hline
\end{tabular}
\caption{Correlations among the lepton universality constraints.}
\label{LeptUnivCorr}
\end{table}

%%%%%%%%%%%%%%%%%%%%%%%%%%%%%%%%%%%%%%%%%%%%%

\subsection{Kaon and $\tau$ decay}

\noindent
Paralleling the treatment of pion decays, we can extract 
$g_\tau/g_\mu$ from kaon decays.  The tree level
decay widths involving kaons are 
\begin{equation}
\Gamma(K\rightarrow \mu\,\bar{\nu}_\mu)
= \frac{g_\mu^2 g_{us}^2}{256\pi}\,\frac{f_K^2}{M_W^4}\,
  m_\mu^2 m_K\left(1-\frac{m_\mu^2}{m_K^2}\right)^2\;,
\end{equation}
and
\begin{equation}
\Gamma(\tau\rightarrow K\,\nu_\tau)
= \frac{g_\tau^2 g_{us}^2}{512\pi}\,\frac{f_K^2}{M_W^4}\,
  m_\tau^3\left(1-\frac{m_K^2}{m_\tau^2}\right)^2\;,
\end{equation}
where $g_{us} = g|V_{us}|$, and the kaon decay constant $f_K$ is normalized
as (Ref.~\cite{Hagiwara:fs}, page 439)
\begin{equation}
\langle 0 |\,\bar{u}\gamma_\mu \gamma_5 s \,(0)\,| K^-(\mathbf{q}) \rangle = i\,q_\mu f_K\;.
\end{equation}
Taking the ratio yields
\begin{equation}
R^0_{\tau/K} \;\equiv\;
\frac{\Gamma(\tau\rightarrow K\,\nu_\tau      )}
     {\Gamma(\pi \rightarrow\mu\,\bar{\nu}_\mu)}
\;=\;
\frac{\tau_K}{\tau_\tau}\,
\frac{B(\tau\rightarrow K \,\nu_\tau   )}
     {B(K \rightarrow\mu\,\bar{\nu}_\mu)}
\;=\;
\frac{g_\tau^2}{g_\mu^2}\,
\frac{m_\tau^3}{2 m_\mu^2 m_K}\;
\frac{(1-m_K^2/m_\tau^2)^2}
     {(1-m_\mu^2  /m_K^2)^2} \;.
\end{equation}
Radiative corrections modify this to
\begin{equation}
R_{\tau/K}
\;=\;
\frac{\tau_K}{\tau_\tau}\,
\frac{B(\tau\rightarrow K \,\nu_\tau    \,(\gamma))}
     {B(K \rightarrow\mu\,\bar{\nu}_\mu \,(\gamma))}
\;=\;
R^0_{\tau/K}
\left( 1 + \delta R_{\tau/K} \right)\;,
\end{equation}
with \cite{Decker:1994ea}
\begin{equation}
\delta R_{\tau/K} = + 0.0090^{+0.0017}_{-0.0026}\;.
\end{equation}
Using this relation and the data listed in 
Table~\ref{MassLifeBranch}, we obtain
\begin{equation}
(g_\tau/g_\mu)_{K\tau} = 0.979\pm 0.017 \;, 
\end{equation}
which agrees with Ref.~\cite{Pich:2002bc}.
This has a correlation of $+0.07$ with 
$(g_\tau/g_e)_{\tau\mu}$ of Eq.~(\ref{taudecay}), 
and a correlation of $+0.04$ with
$(g_\tau/g_\mu)_{\pi\tau}$ of Eq.~(\ref{pidecay}),
arising from the common inputs $\tau_\tau$ and $m_\tau$.

We tabulate our results in Tables~\ref{LeptUnivConst} and
\ref{LeptUnivCorr}.

%%%%%%%%%%%%%%%%%%%%%%%%%%%%%%%%%%%%%%%%%%%%%

\section{$Z$-pole, NuTeV, and $W$ mass data}

For the $Z$-pole and NuTeV data,
we use the same set as in Ref.~\cite{LOTW1}, namely
$\Gamma_\mathrm{lept}$, $\Gamma_\mathrm{inv}/\Gamma_\mathrm{lept}$,
and $\sin^2\theta_\mathrm{eff}^\mathrm{lept}$ from
$e^+e^-$ colliders, $g_L^2$ and $g_R^2$ from NuTeV.
Of these, only the value of $\sin^2\theta_\mathrm{eff}^\mathrm{lept}$
has been updated since our last analysis in Ref.~\cite{LOTW1}.
The $W$ mass has also been updated by LEP-II.
We list the values in Table~\ref{ZpoleData}.
There is a correlation of $0.17$ between $\Gamma_\mathrm{lept}$ and
$\Gamma_\mathrm{inv}/\Gamma_\mathrm{lept}$;
other correlations are negligible.

\begin{table}
\begin{tabular}{c|c|c|l}
\,Observable\, &  \,SM prediction\,  &  \,Measured Value\, & \,Reference \\
\hline
$M_Z$ & Input & $\;91.1875\pm 0.0021$~GeV\, & \ \cite{LEP/SLD:2003} page 8 \\
$\Gamma_\mathrm{lept}$ &
$84.034$~MeV &
\ $83.984\pm 0.086$~MeV\  & 
\ \cite{LEP/SLD:2003} page 9 \\
$\Gamma_\mathrm{inv}/\Gamma_\mathrm{lept}$ &
$5.972$ &
$5.942 \pm 0.016$ &
\ \cite{LEP/SLD:2003} page 8 \\
$\sin^2\theta_\mathrm{eff}^\mathrm{lept}$ &
$0.23133$ &
$0.23150\pm 0.00016$ &
\ \cite{LEP/SLD:2003} page 142 \\
$g_L^2$ &
$0.3040$ &
$0.3002 \pm 0.0012$ &
\ Average of \cite{Zeller:2001hh} and \cite{PDG1998} \\
$g_R^2$ &
$0.0304$ &
$0.0310 \pm 0.0010$ &
\ Average of \cite{Zeller:2001hh} and \cite{PDG1998} \\
$M_W$ &
$80.399$~GeV &
\ $80.426\pm 0.034$~GeV\  &
\ \cite{LEP/SLD:2003} page 146 \\
\hline
\end{tabular}
\caption{The observables used in this analysis
in addition to the lepton universality data.
The measured value of $\sin^2\theta_\mathrm{eff}^\mathrm{lept}$
and the $W$ mass have been updated in Ref.~\cite{LEP/SLD:2003}
since the analysis in Ref.~\cite{LOTW1}.
The SM predictions are ZFITTER \cite{ZFITTER} 
outputs with inputs of
$M_\mathrm{top}=178.0\,\mathrm{GeV}$ \cite{Azzi:2004rc},
$M_\mathrm{Higgs}=115\,\mathrm{GeV}$,
$\alpha_s(M_Z) = 0.119$, and
$\Delta\alpha_\mathrm{had}^{(5)} = 0.02755$ \cite{Hagiwara:2003da}.}
\label{ZpoleData}
\end{table}

%%%%%%%%%%%%%%%%%%%%%%%%%%%%%%%%%%%%%%%%%%%%%%%%%%%%%%%%%%%%%%%%%%%%%%%
\section{The Corrections}

Suppression of the neutrino-gauge couplings modifies the relation
between the Fermi constant $G_F$ and the muon decay constant $G_\mu$ to
\begin{equation}
G_F = G_\mu\left( 1 + \dfrac{\varepsilon_e + \varepsilon_\mu}{2}
           \right)\;.
\label{FermiConstant}
\end{equation}
Since $G_\mu$ is used as an input in calculating the SM predictions, 
all observables whose SM prediction depends on $G_F$ will
receive this correction through $G_F$.
Of the observables that measure the $Z\nu_\ell\nu_\ell$
and $W\ell\nu_\ell$ vertices directly,
the $Z$ invisible width is corrected by an additional factor of 
\begin{equation}
1-\frac{2}{3}\left(\varepsilon_e + \varepsilon_\mu + \varepsilon_\tau
             \right) \;,
\end{equation}
while the NuTeV parameters $g_L^2$ and $g_R^2$ receive an
additional correction of $(1-\varepsilon_\mu)$.
The dependence of the observables on the oblique correction parameters
$S$, $T$, and $U$ can be found elsewhere \cite{Peskin:1990zt}.

Numerically, the observables are corrected as follows:
\begin{eqnarray}
\dfrac{  \Gamma_\mathrm{lept} }
      { [\Gamma_\mathrm{lept}]_\mathrm{SM} }
& = & 1 - 0.0021\,S + 0.0093\,T 
      + 0.60\,\varepsilon_e + 0.60\,\varepsilon_\mu\;, \cr
\dfrac{  \Gamma_\mathrm{inv}/\Gamma_\mathrm{lept} }
      { [\Gamma_\mathrm{inv}/\Gamma_\mathrm{lept}]_\mathrm{SM} }
& = & 1 + 0.0021\,S - 0.0015\,T
      - 0.76\,\varepsilon_e - 0.76\,\varepsilon_\mu - 0.67\,\varepsilon_\tau\;, \cr
\dfrac{  \sin^2\theta_{\mathrm{eff}}^{\mathrm{lept}} }
      { [\sin^2\theta_{\mathrm{eff}}^{\mathrm{lept}}]_\mathrm{SM} }
& = & 1 + 0.016\,S - 0.011\,T
      - 0.72\,\varepsilon_e - 0.72\,\varepsilon_\mu \;, \cr
\dfrac{  g_L^2 }
      { [g_L^2]_\mathrm{SM} }
& = & 1 - 0.0090\,S + 0.022\,T
      + 0.41\,\varepsilon_e - 0.59\,\varepsilon_\mu \;, \cr
\dfrac{  g_R^2 }
      { [g_R^2]_\mathrm{SM} }
& = & 1 + 0.031\,S - 0.0067\,T
      - 1.4\,\varepsilon_e -2.4\,\varepsilon_\mu \;, \cr
\dfrac{  M_W }
      { [M_W]_\mathrm{SM} }
& = & 1 - 0.0036\,S + 0.0056\,T + 0.0042\,U 
%\cr &   & \qquad 
+ 0.11\,\varepsilon_e + 0.11\,\varepsilon_\mu \;.
\label{STUeps}
\end{eqnarray}
Here, $[*]_\mathrm{SM}$ is the usual SM prediction of the 
observable $*$ using $G_\mu$ as input.

Despite the fact that six observables are available to the fit, this set of
data is not sufficient to fix all six parameters.
This is because the ratio $g_L^2/g_R^2$ depends on the fit
parameters only through $\sin^2\theta_{\mathrm{eff}}^{\mathrm{lept}}$,
and thus can only constrain the exact same linear combination of 
$S$, $T$, $\varepsilon_e$, and $\varepsilon_\mu$ as
$\sin^2\theta_{\mathrm{eff}}^{\mathrm{lept}}$.
In fitting the parameters to the observables, the linear combination
\begin{equation}
\alpha T + 2( \varepsilon_\mu - 2\varepsilon_e) \;,
\end{equation}
remains unconstrained. 
Therefore, we can constrain only five of the six fit
parameters with the $Z$-pole and NuTeV data.

The linear combinations constrained by the
lepton universality bounds from $W$, $\tau$, $\pi$, and $K$-decays are
\begin{eqnarray}
\frac{g_\mu}{g_e}    & = & 1 + \frac{\varepsilon_e   - \varepsilon_\mu }{2} \;, \cr
\frac{g_\tau}{g_\mu} & = & 1 + \frac{\varepsilon_\mu - \varepsilon_\tau}{2} \;, \cr
\frac{g_\tau}{g_e}   & = & 1 + \frac{\varepsilon_e   - \varepsilon_\tau}{2} \:.
\label{couplingratios}
\end{eqnarray}
Since there are only two independent observables but three fit parameters,
only two independent linear combinations can be simultaneously constrained by the lepton 
universality data.
However, when these bounds are combined with the $Z$-pole and NuTeV data, 
all three $\varepsilon_\ell$ can be constrained independently.

%%%%%%%%%%%%%%%%%%%%%%%%%%%%%%%%%%%

\section{Fits to Lepton Universality Constraints}

\begin{figure}[p]
\begin{center}
\raisebox{6.2cm}{(a)}
\rotatebox{90}{\scalebox{0.4}{\includegraphics{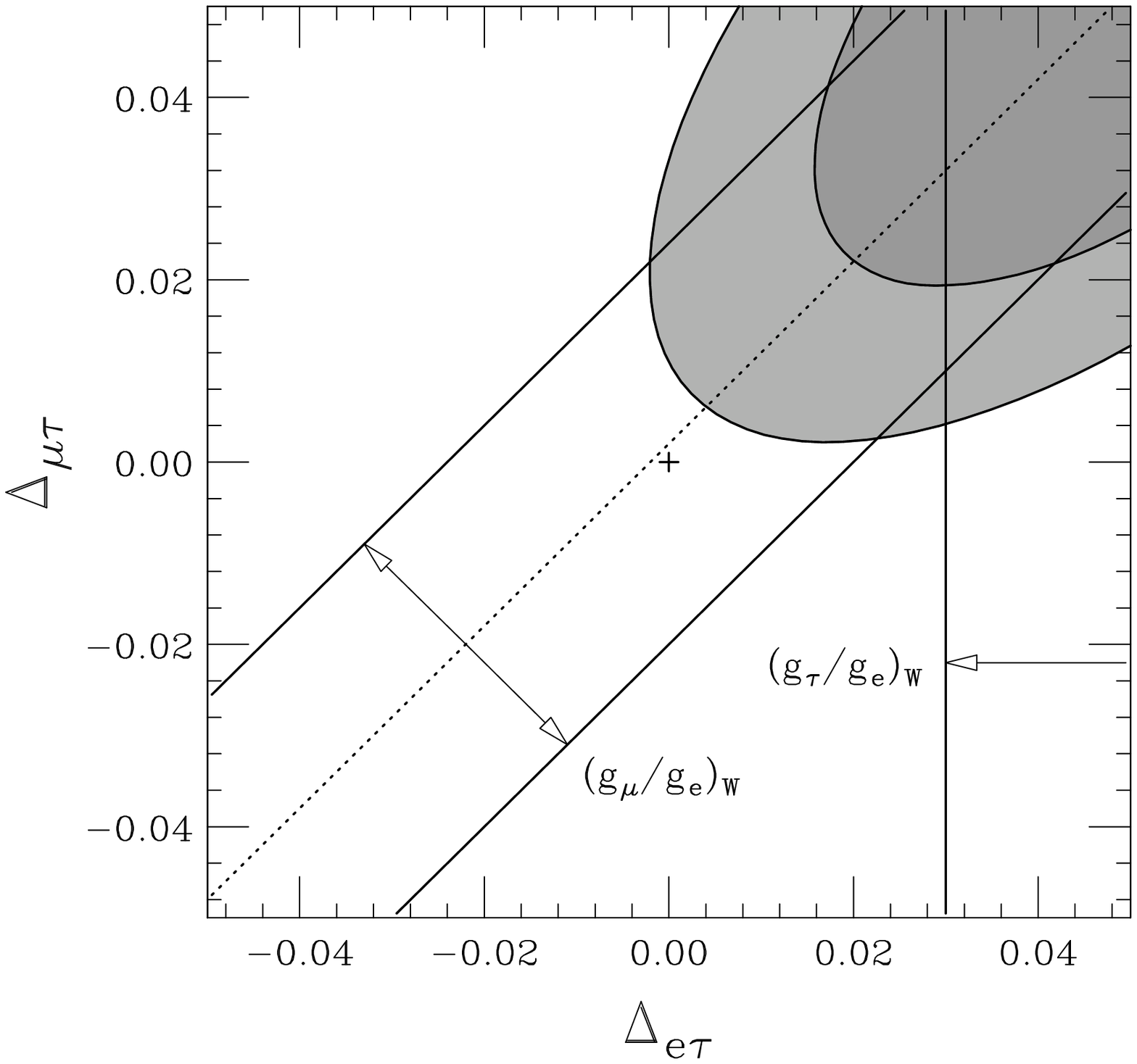}}}\quad
\raisebox{6.2cm}{(b)}
\rotatebox{90}{\scalebox{0.4}{\includegraphics{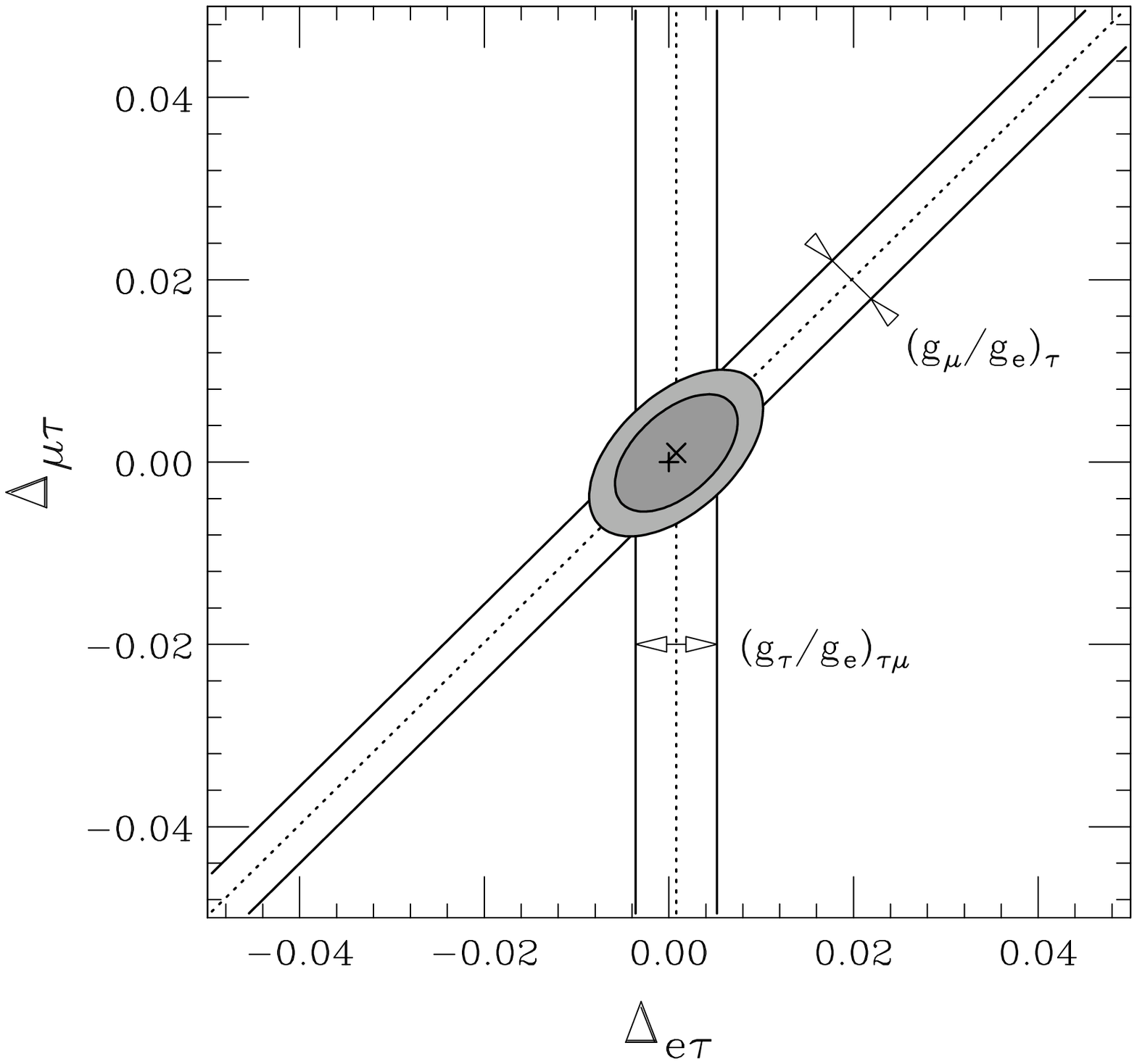}}}\\
\raisebox{6.2cm}{(c)}
\rotatebox{90}{\scalebox{0.4}{\includegraphics{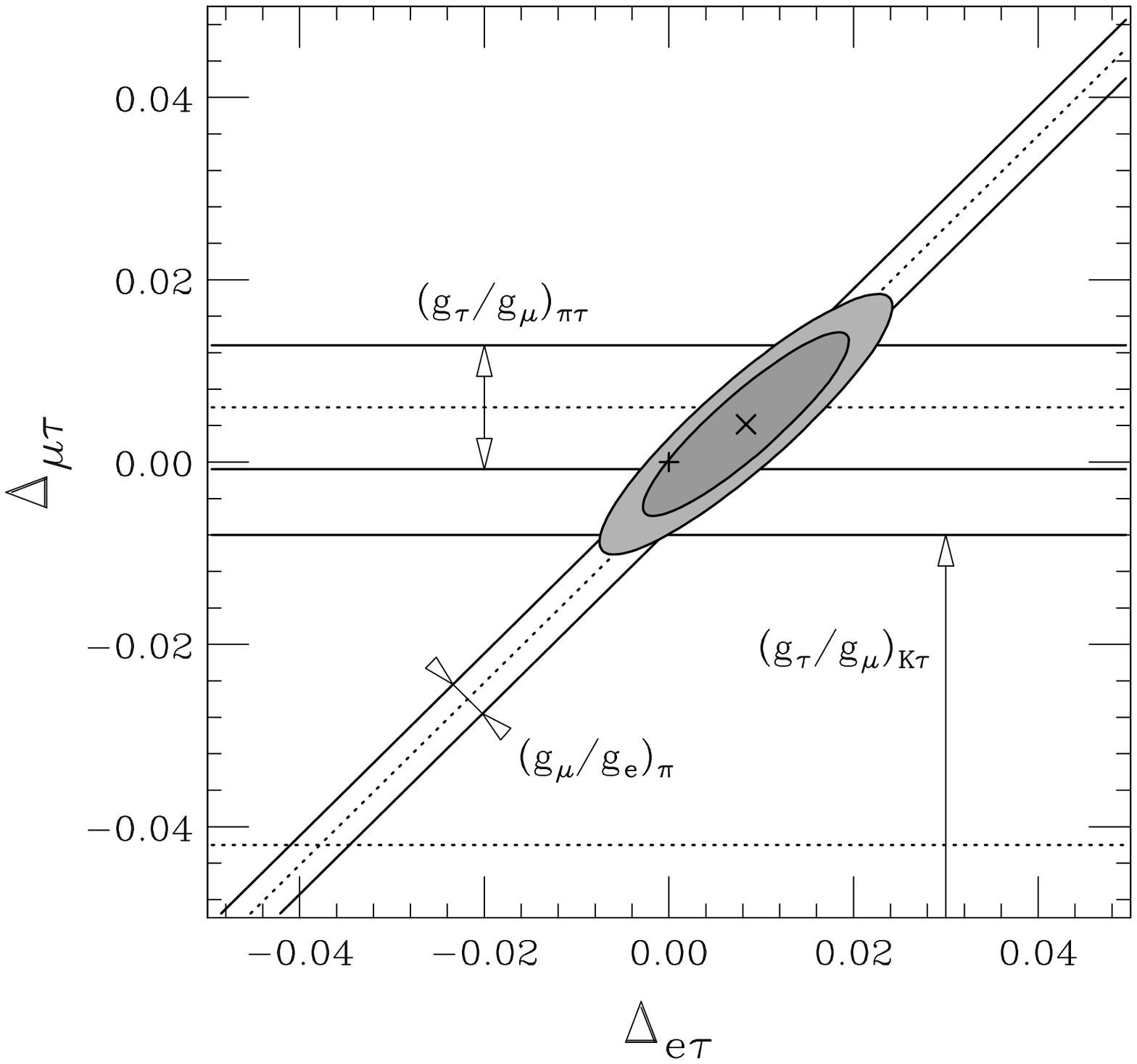}}}\quad
\raisebox{6.2cm}{(d)}
\rotatebox{90}{\scalebox{0.4}{\includegraphics{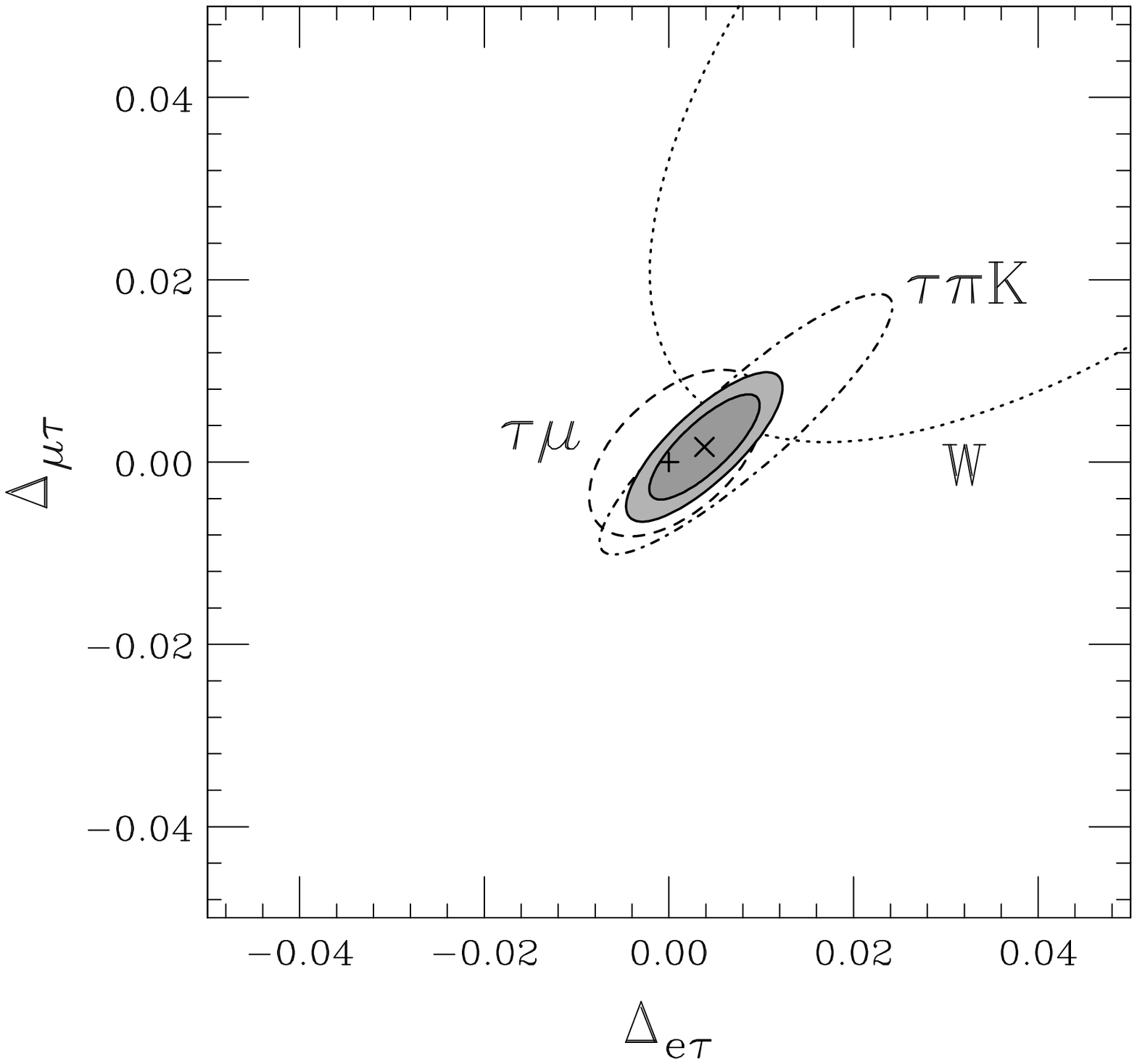}}}
\caption{The limits on $\Delta_{e\tau}$ and $\Delta_{\mu\tau}$ from
(a) $W$-decay, (b) $\tau$-decay, (c) $\pi$ and $K$-decay, 
and (d) all decays combined.
The $1\sigma$ bands are shown for each coupling constant ratio
ignoring correlations.
The shaded areas represent the 68\% (dark gray) and 90\% (light gray)
confidence contours including correlations.}
\label{emufig}
\end{center}
\end{figure}

The three ratios of the coupling constants contain only two independent degrees of freedom, 
since the third is the product of the first two. 
Equivalently, when parameterizing the ratios in terms of the differences of the $\varepsilon_\ell$, 
as in Eq.~(\ref{couplingratios}), only two (\textit{any} two) sets of the differences 
can be taken as free parameters; the third can be expressed as the difference between them.  
We can combine the seven pieces of experimental data in Table~\ref{LeptUnivConst} 
by fitting with any two of the three parameters
\begin{eqnarray}
\Delta_{e\mu}    
& \equiv & \varepsilon_e  - \varepsilon_\mu 
\;=\; \Delta_{e\tau} + \Delta_{\mu\tau} \;,\cr
\Delta_{\mu\tau} 
& \equiv & \varepsilon_\mu  - \varepsilon_\tau 
\;=\; \Delta_{e\tau} - \Delta_{e\mu}    \;,\cr
\Delta_{e\tau}   
& \equiv & \varepsilon_e  - \varepsilon_\tau 
\;=\; \Delta_{e\mu} - \Delta_{\mu\tau} \;. 
\label{Deltadefs}
\end{eqnarray}
We obtain
\begin{eqnarray}
\Delta_{e\mu}    & = & 0.0022 \pm  0.0025 \;, \cr
\Delta_{\mu\tau} & = & 0.0017 \pm  0.0038 \;, \cr
\Delta_{e\tau}   & = & 0.0039 \pm  0.0040 \;,
\end{eqnarray}
with correlations shown in Table~\ref{DeltaCorr}.
In terms of the coupling constant ratios, this translates to
\begin{eqnarray}
(g_\mu/g_e)  & = & 1.0011 \pm 0.0012 \cr
(g_\tau/g_e) & = & 1.0019 \pm 0.0020
\label{LeptUnivAverage}
\end{eqnarray}
with a correlation of $0.37$.
The quality of the fit is unimpressive:  the $\chi^2$ is 8.4 
for (7-2)=5 degrees of freedom. 
The largest contribution is from
$(g_\tau/g_e)_W$ which contributes $4.6$.
The region of $\Delta_{e\tau}$-$\Delta_{\mu\tau}$ parameter space preferred by the fit is shown
in Fig.~\ref{emufig}.  The 90\% confidence contour
preferred by the $W$-decay data hardly overlaps with that of the
$\tau$-decay data, which causes the large $\chi^2$.  

Since the objective of this paper is to determine whether the $Z$-pole and NuTeV data are
compatible with lepton non-universality, it is problematic that the 
lepton universality constraints are not clearly compatible among themselves.
The set of coupling ratios we consider here has an intrinsic $\chi^2$ of 8.4 (or 10.8, if the 
ALEPH value is excluded from the calculation of $B(\tau\rightarrow \pi\nu_\tau)$)
which cannot be mitigated in our model.  
This is in addition to the large $\chi^2$ associated with $B(\tau\rightarrow \pi\nu_\tau)$, 
discussed previously.  
Further experiments may provide the ultimate resolution of the tension in the data.
For now, to prevent this large $\chi^2$ among the
lepton universality data from obscuring their compatibility with the $Z$-pole and NuTeV
data, we will use the average values obtained in Eq.~(\ref{LeptUnivAverage}) 
in our subsequent fits with the caveat that the pair hides a large $\chi^2$.

\begin{table}[t]
\begin{tabular}{l|rrr}
& $\Delta_{e\mu}$ 
& $\Delta_{\mu\tau}$ 
& $\Delta_{e\tau}$ \\
\hline
$\Delta_{e\mu}$    & $\phantom{-}1.00$ & $-0.27$ & $0.37$ \\
$\Delta_{\mu\tau}$ &                   &  $1.00$ & $0.80$ \\
$\Delta_{e\tau}$   &                   &         & $\phantom{-}1.00$ \\
\hline
\end{tabular}
\caption{Correlations among the $\Delta$'s from fit.}
\label{DeltaCorr}
\end{table}

%%%%%%%%%%%%%%%%%%%%%%%%%%%%%%%%%%%%%%%%%%%%%%%%%%%%%%%%%%%%%%%%%%%%%%

\section{Fits to $Z$-pole, NuTeV, and Lepton Universality data}

We fit the expressions in section~IV to the $Z$-pole, NuTeV, 
and $W$ mass data listed in Table~\ref{ZpoleData}, 
and the lepton universality constraint Eq.~(\ref{LeptUnivAverage}).
The $S$, $T$, $U$ parameters were used in all fits.
Of the three $\varepsilon_\ell$ we performed fits with the following eight combinations of fit parameters:
\begin{enumerate}[A.]
\item fit with a flavor-universal $\varepsilon$ ($\varepsilon_e=\varepsilon_\mu= \varepsilon_\tau=\varepsilon$),
\item fit with all three parameters $\varepsilon_e$, $\varepsilon_\mu$, and $\varepsilon_\tau$,
\item fit with $\varepsilon_e$ and $\varepsilon_\mu$,
\item fit with $\varepsilon_e$ and $\varepsilon_\tau$,
\item fit with $\varepsilon_\mu$ and $\varepsilon_\tau$,
\item fit with $\varepsilon_e$ only,
\item fit with $\varepsilon_\mu$ only,
\item fit with $\varepsilon_\tau$ only.
\end{enumerate}
Fit A with flavor-universal $\varepsilon$ is the one performed in 
Ref.~\cite{LOTW1} (without the lepton universality constraints).
We include it here as a benchmark against which to compare 
the flavor-dependent fits.
The reference Standard Model values were calculated using ZFITTER \cite{ZFITTER}
with the inputs $M_Z=91.1875~\mathrm{GeV}$ \cite{LEP/SLD:2003}, $M_H=115~\mathrm{GeV}$, 
$M_t=178.0~\mathrm{GeV}$ \cite{Azzi:2004rc}, $\alpha_s(M_Z)=0.119$, and
$\Delta\alpha^{(5)}_\mathrm{had}(M_Z) = 0.02755$ \cite{Hagiwara:2003da}.

The results of these fits have been tabulated in Table~\ref{AllFitResults},
with correlations among the fit parameters shown in 
Table~\ref{epsCorr}.
As the values of $\chi^2$ in Table~\ref{AllFitResults} indicate, 
the quality of the fits A, B, C, D, and F is excellent, 
while fits E and G are only marginal and fit H fails.
The largest contribution to the overall $\chi^2$ for fits
E and G is from $(g_\mu/g_e)$ (5.1 for E and 5.3 for G)
which indicates that these fits are not compatible with lepton
universality.
For fit H, the largest contributions to the overall $\chi^2$ is
from the NuTeV observable $g_L^2$ (5.5) and $(g_\tau/g_e)$ (3.0)
which indicates that neither NuTeV nor lepton universality are
accommodated.
Comparisons of fits B and C, D and F, E and G show that including
$\epsilon_\tau$ in the fits does little to improve the overall $\chi^2$.
Indeed, the $\chi^2$ per degree of freedom in actually worse with the
inclusion of $\epsilon_\tau$.

%%%%%%%%%%%%%%%%%%%%%%%%%%%%%%%%%%%%%%%%%%%%%%%%%%%%%%%%%%%%%%%%%%%%%%%%%%%%%

\begin{table}[t]
\begin{tabular}{|c||c|c|c|c|}
\hline
fit parameters & A & B & C & D \\
\hline
$S$ &
$-0.01\pm 0.10$ &
$\phantom{-}0.00\pm 0.10$ &
$\phantom{-}0.00\pm 0.10$ &
$-0.04\pm 0.10$ \\
\hline
$T$ &
$-0.48\pm 0.15$ &
$-0.56\pm 0.16$ &
$-0.56\pm 0.16$ &
$-0.45\pm 0.15$ \\
\hline
$U$ &
$\phantom{-}0.55\pm 0.16$ &
$\phantom{-}0.62\pm 0.17$ &
$\phantom{-}0.62\pm 0.17$ &
$\phantom{-}0.51\pm 0.16$ \\
\hline
$\varepsilon_e$ &
$\;0.0030\pm 0.0010\;$ &
$\;0.0048\pm 0.0018\;$ &
$\;0.0049\pm 0.0018\;$ &
$\;0.0050\pm 0.0018\;$ \\
\hline
$\varepsilon_\mu$ &
$\;0.0030\pm 0.0010\;$ &
$\;0.0027\pm 0.0014\;$ &
$\;0.0027\pm 0.0014\;$ &
--- \\
\hline
$\varepsilon_\tau$ &
$\;0.0030\pm 0.0010\;$ &
$\;0.0007\pm 0.0028\;$ &
--- &
$\;0.0012\pm 0.0028\;$ \\
\hline
\hline
$\chi^2$ & $2.4$ & $0.91$ & $0.97$ & $4.4$ \\
\hline
d.o.f.   & 4 & 2 & 3 & 3 \\
\hline
$\chi^2/\mathrm{d.o.f.}$ & 
$0.6$ & $0.5$ & $0.3$ & $1.5$ \\
\hline
large $\chi^2$
& --- & --- & --- & --- \\
&     &     &     &     \\
\hline
\end{tabular}

\vspace{0.5cm}

\begin{tabular}{|c||c|c|c|c|}
\hline
fit parameters & E & F & G & H \\
\hline
$S$ &
$-0.03\pm 0.10$ &
$-0.04\pm 0.10$ &
$-0.03\pm 0.10$ &
$-0.08\pm 0.10$ \\
\hline
$T$ &
$-0.30\pm 0.13$ &
$-0.46\pm 0.15$ &
$-0.30\pm 0.13$ &
$-0.18\pm 0.12$ \\
\hline
$U$ &
$\phantom{-}0.38\pm 0.15$ &
$\phantom{-}0.52\pm 0.16$ &
$\phantom{-}0.37\pm 0.15$ &
$\phantom{-}0.25\pm 0.13$ \\
\hline
$\varepsilon_e$ &
--- &
$\;0.0051\pm 0.0018\;$ &
--- &
--- \\
\hline
$\varepsilon_\mu$ &
$\;0.0028\pm 0.0014\;$ &
--- &
$\;0.0029\pm 0.0014\;$ &
--- \\
\hline
$\varepsilon_\tau$ &
$\;0.0021\pm 0.0027\;$ &
--- &
--- &
$\;0.0026\pm 0.0027\;$ \\
\hline
\hline
$\chi^2$ & $7.8$ & $4.5$ & $8.3$ & $11.6$ \\
\hline
d.o.f.   & 3 & 4 & 4 & 4 \\
\hline
$\chi^2/\mathrm{d.o.f.}$ & 
$2.6$ & $1.1$ & $2.1$ & $2.9$ \\
\hline
large $\chi^2$
& $5.1$ from $(g_\mu/g_e)$  
& --- 
& $5.3$ from $(g_\mu/g_e)$ 
& $5.5$ from $g_L^2$ \\
& $2.5$ from $(g_\tau/g_e)$
& & & $3.0$ from $(g_\tau/g_e)$ \\
\hline
\end{tabular}
\caption{The results of the fits.}
\label{AllFitResults}
\end{table}

%%%%%%%%%%%%%%%%%%%%%%%%%%%%%%%%%%%%%%%%%%%%%%%%%%%%%%%%%%%%%%%%%%%%%%%%%%%%%

\begin{figure}[p]
\begin{center}
\raisebox{6.2cm}{(a)}
\rotatebox{90}{\scalebox{0.4}{\includegraphics{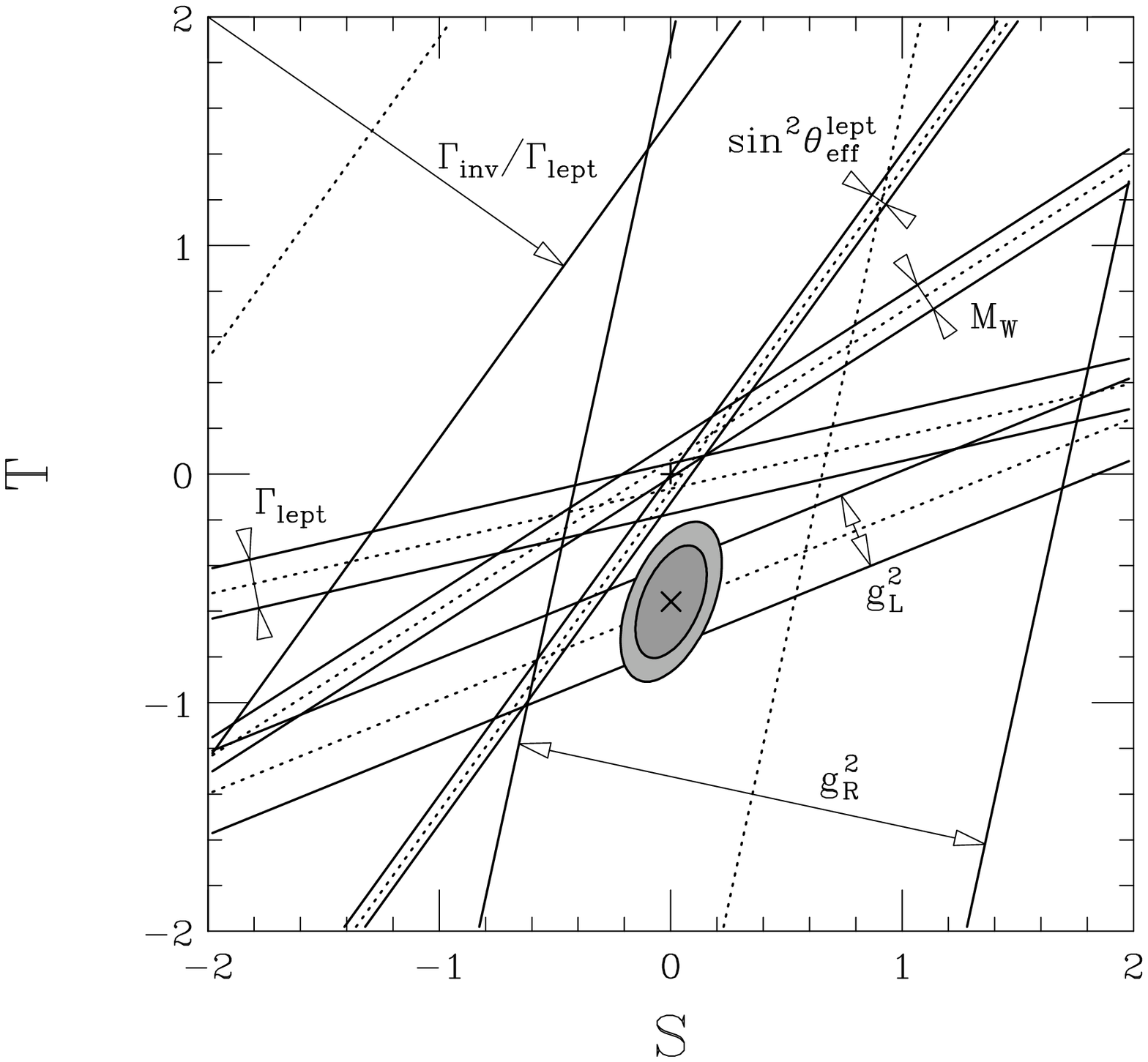}}}\quad
\raisebox{6.2cm}{(d)}
\rotatebox{90}{\scalebox{0.4}{\includegraphics{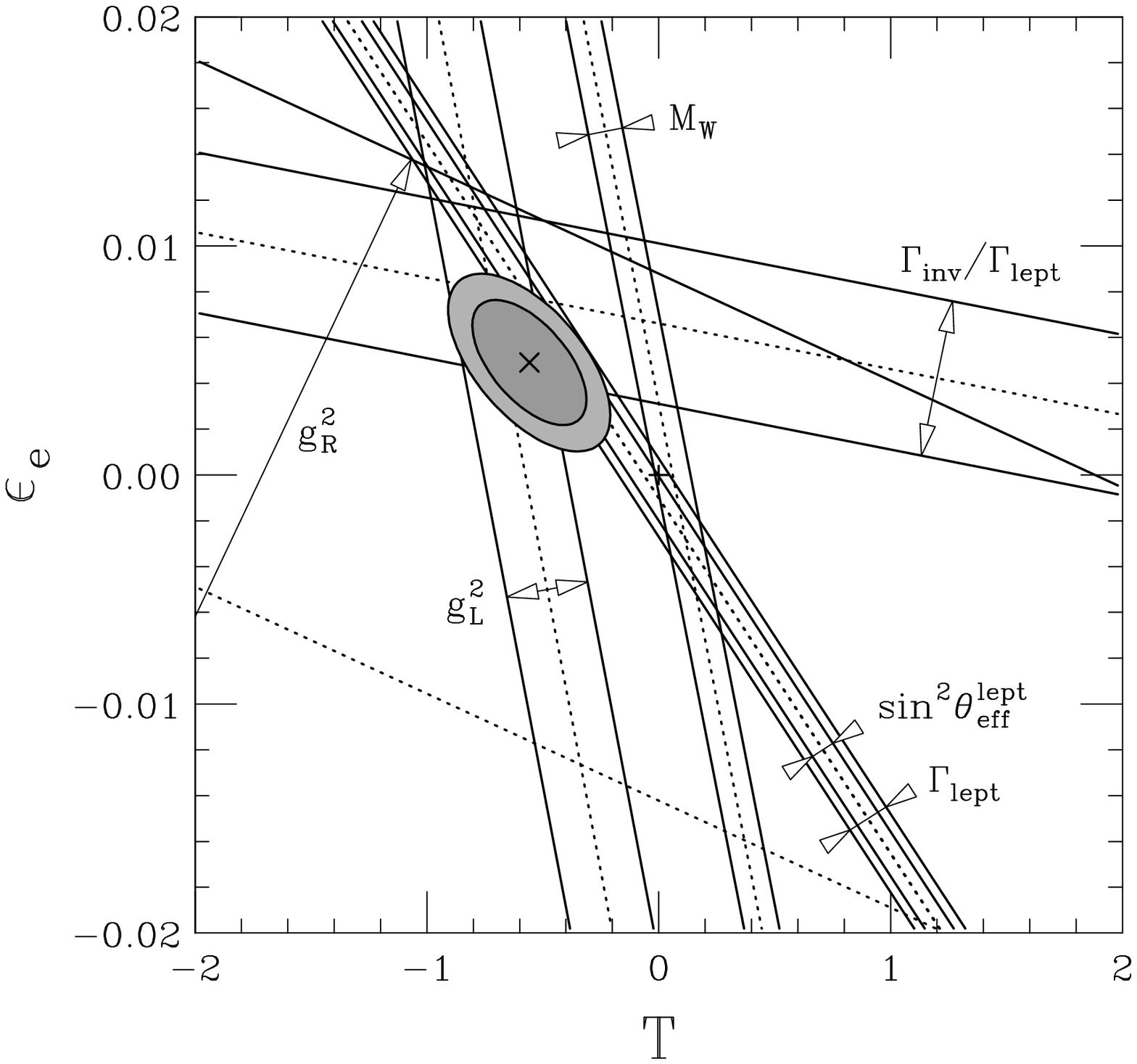}}}\\
\raisebox{6.2cm}{(b)}
\rotatebox{90}{\scalebox{0.4}{\includegraphics{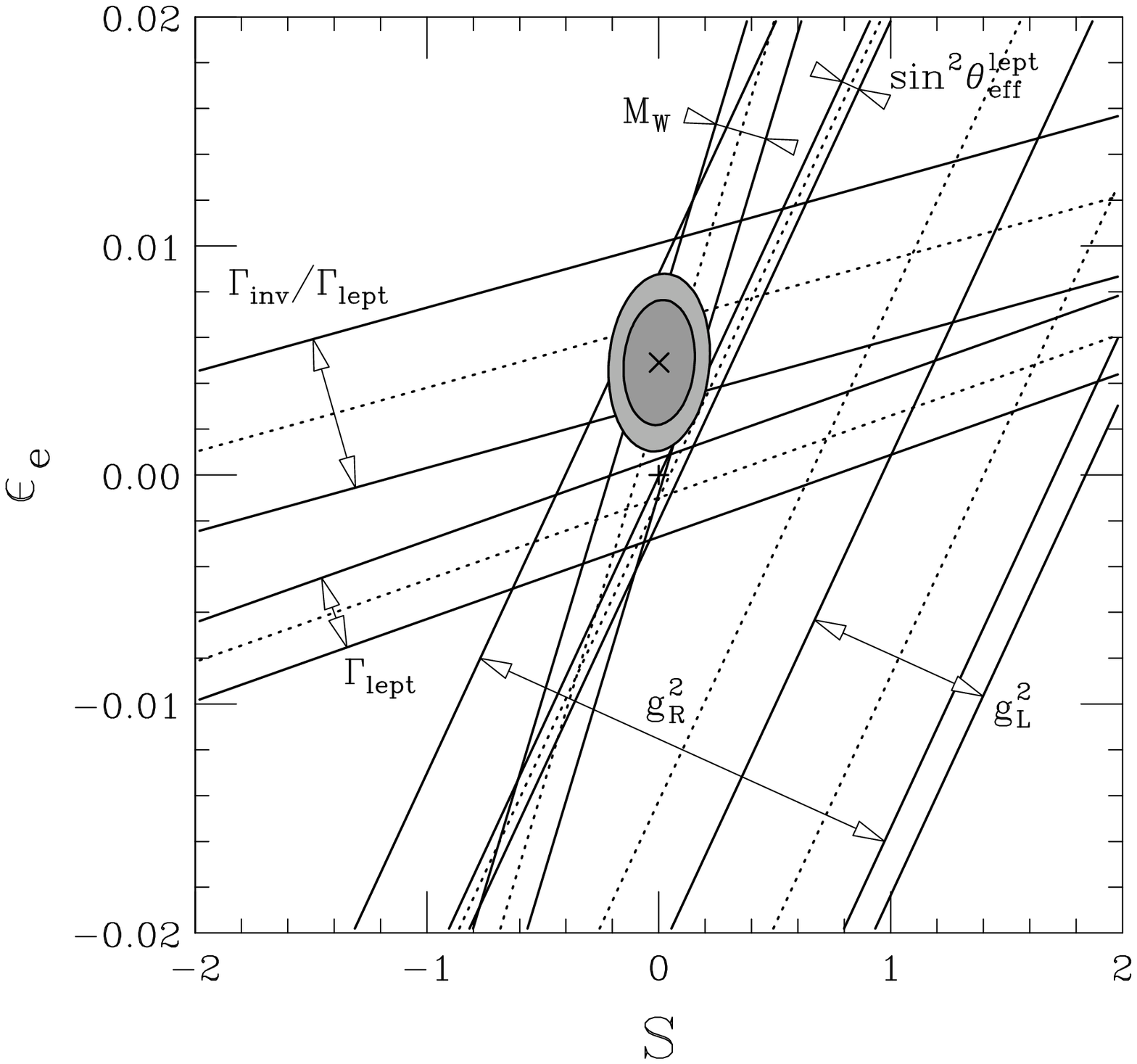}}}\quad
\raisebox{6.2cm}{(e)}
\rotatebox{90}{\scalebox{0.4}{\includegraphics{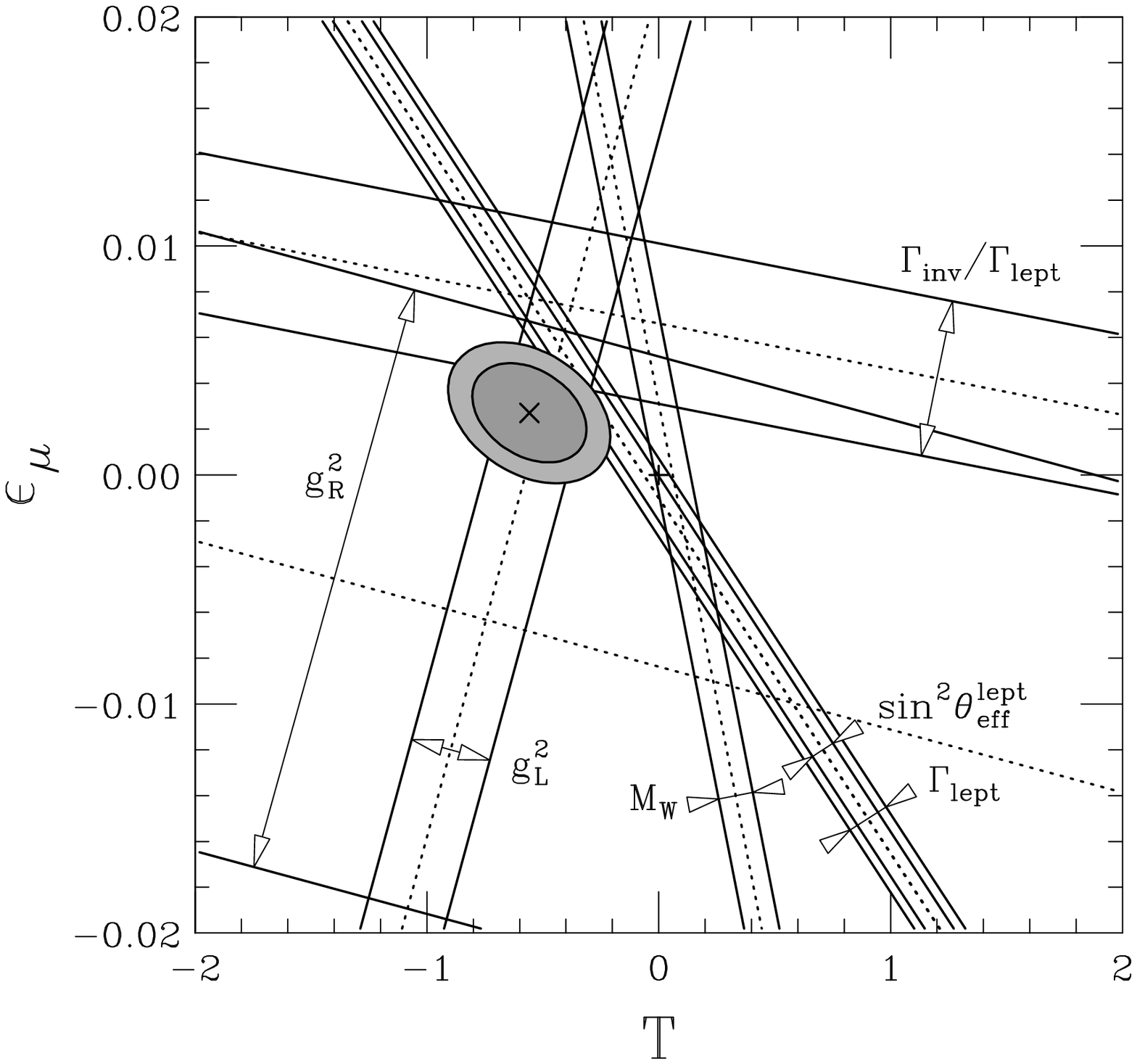}}}\\
\raisebox{6.2cm}{(c)}
\rotatebox{90}{\scalebox{0.4}{\includegraphics{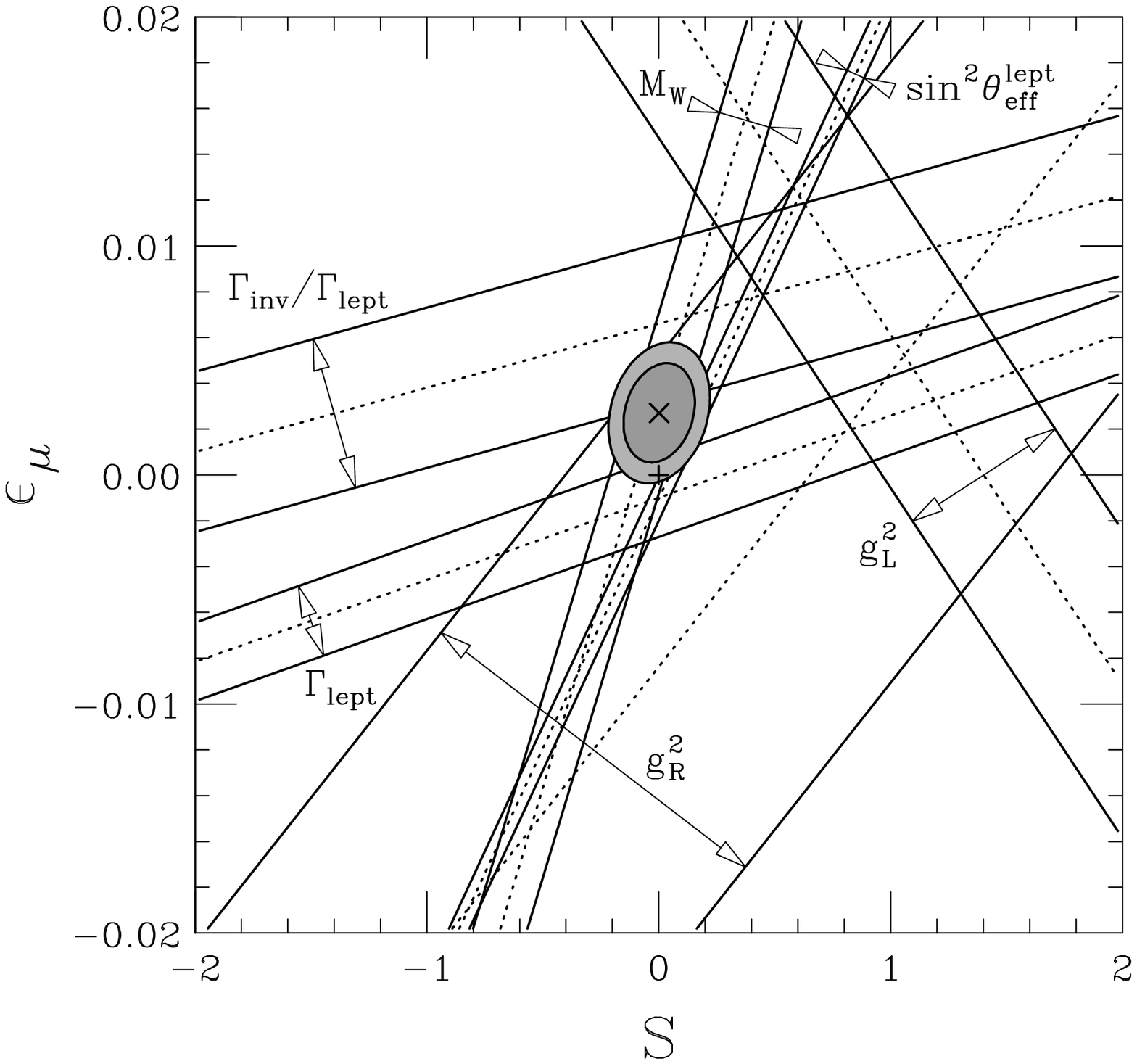}}}\quad
\raisebox{6.2cm}{(f)}
\rotatebox{90}{\scalebox{0.4}{\includegraphics{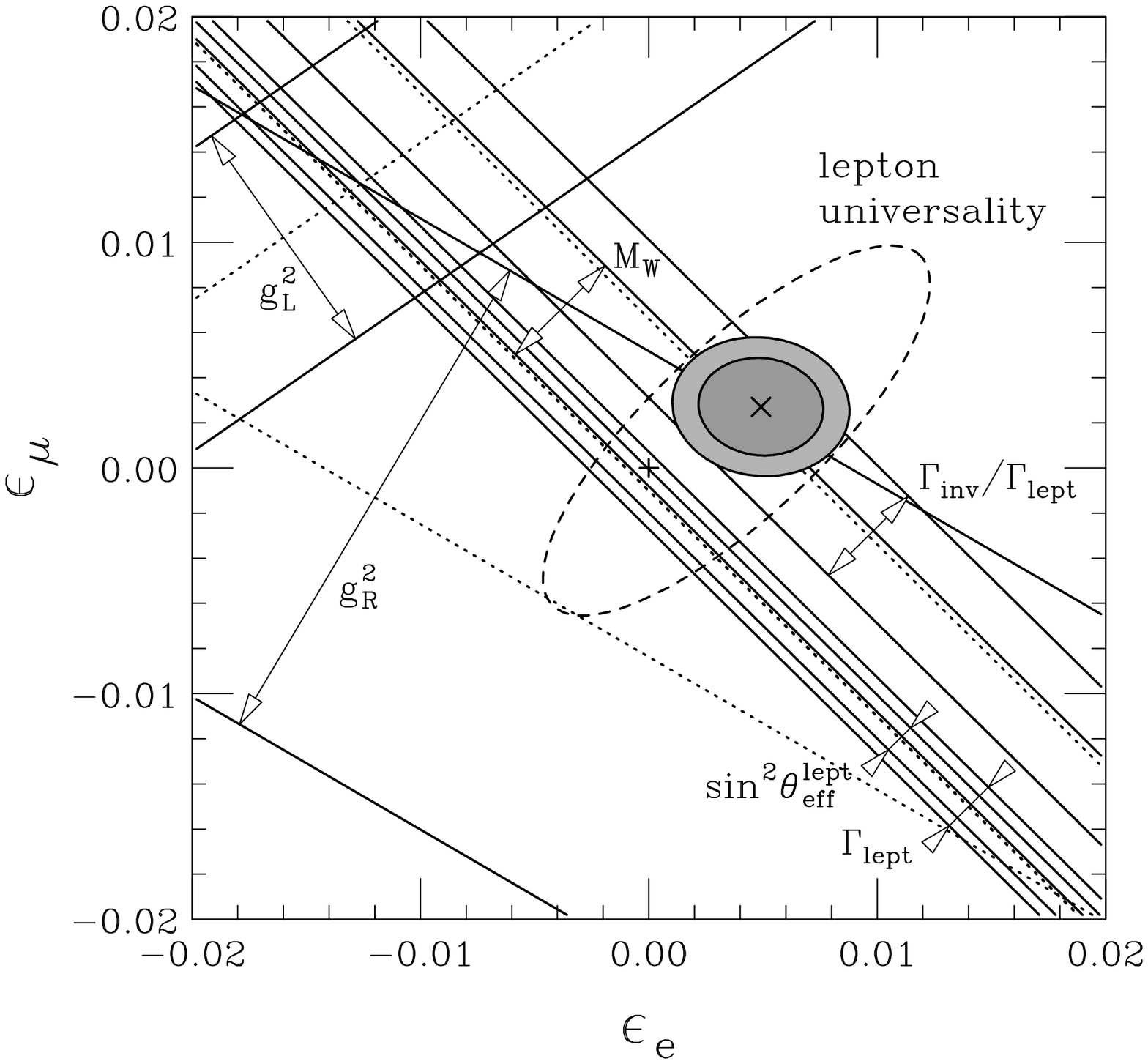}}}
\caption{The 68\% and 90\% confidence contours projected onto various planes for 
the five-paremeter fit with $S$, $T$, $U$, $\varepsilon_e$ and $\varepsilon_\mu$.
The bands associated with each observable show the $1\sigma$ limits in the 
respective planes.
The origin is the reference SM with $M_\mathrm{top}=178.0\,\mathrm{GeV}$ and
$M_\mathrm{Higgs}=115\,\mathrm{GeV}$. The dashed contour in 2(f) is the 90\%
lepton universality bound.}
\label{fitCfig}
\end{center}
\end{figure}

The constraints placed on the fit parameters by each observable are
illustrated in figure~\ref{fitCfig}.
The 1$\sigma$ bands in each 2 dimensional plane are plotted assuming that 
all other fit parameters are set to zero.
The gray ellipses are the 68\% and 90\% confidence regions for fit C,
\textit{i.e.} the five parameter fit with $S$, $T$, $U$, $\varepsilon_e$, 
and $\varepsilon_\mu$,
projected to each plane from the full five-dimensional parameter space.  
In the case of the $M_W$ band, since $M_W$ serves only to fix $U$, 
it exerts no statistical `pull' on the other fit parameters; 
also, since $U$ is fixed by a single observable we have not included 
figures with $U$ on an axis.  
We have also omitted projections onto planes involving $\varepsilon_\tau$
since $\varepsilon_\tau$
serves little in improving the quality of the fits
and since, 
other than the lepton universality constraint on $(g_\tau/g_e)$, 
it is constrained by only $\Gamma_\mathrm{inv}/\Gamma_\mathrm{lept}$.

Figure~\ref{fitCfig} clarifies the reason for the failure of fit H.
From figure~\ref{fitCfig}(a), we see that the NuTeV observable $g_L^2$ prefers a
negative $T$.  
To maintain the agreement between the
SM predictions and the $Z$-pole data, the effect of negative $T$ must be absorbed
by a corresponding shift in $G_F$, Eq.~(\ref{FermiConstant}), by making 
$\varepsilon_e$ and/or $\varepsilon_\mu$ positive (as indicated in
figures~\ref{fitCfig}(d) and \ref{fitCfig}(e)).
However, since $\varepsilon_e$ and $\varepsilon_\mu$ are both constrained to zero
in fit H, it cannot accommodate $g_L^2$.  Further, the measured value of 
$\Gamma_\mathrm{inv}/\Gamma_\mathrm{lept}$ is smaller than the
SM prediction, which demands positive $\varepsilon_\tau$, while $(g_\tau/g_e)$, 
Eq.~(\ref{LeptUnivAverage}), prefers negative $\varepsilon_\tau$. 
Thus, H cannot satisfy $(g_\tau/g_e)$ either.

For fits E and G, in which $\varepsilon_e$ is constrained to zero,
the effect of a negative $T$ is absorbed by a positive $\varepsilon_\mu$.
However, the experimental value of $(g_\mu/g_e)$ prefers
$\varepsilon_\mu$ negative.  A tension thus remains between the electroweak
data and the lepton universality data in these fits.

%%%%%%%%%%%%%%%%%%%%%%%%%%%%%%%%%%%%%%%%%%%%%%%%%%%%%%%%%%%%%%%%%%%%%%%%%%%%%

\begin{table}[t]
\begin{tabular}{|c||c|c|}
\hline
fit parameters & E' & F' \\
\hline
$S$ &
$-0.03\pm 0.10$ &
$\phantom{-}0.00\pm 0.10$ \\
\hline
$T$ &
$-0.50\pm 0.17$ &
$-0.38\pm 0.14$ \\
\hline
$U$ &
$\phantom{-}0.56\pm 0.18$ &
$\phantom{-}0.45\pm 0.15$ \\
\hline
$\varepsilon_e$ &
$\;0.0058\pm 0.0023\;$ &
--- \\
\hline
$\varepsilon_\mu$ &
--- &
$\;0.0048\pm 0.0017\;$ \\
\hline
\hline
$\chi^2$ & $4.7$ & $2.9$ \\
\hline
d.o.f.   & 4 & 4 \\
\hline
$\chi^2/\mathrm{d.o.f.}$ & 
$1.2$ & $0.74$ \\
\hline
\hline
\end{tabular}
\caption{The results of the fits using the lepton universality constraint from $\mu$ and $\tau$-decay only.}
\label{TauFitResults}
\end{table}

%%%%%%%%%%%%%%%%%%%%%%%%%%%%%%%%%%%%%%%%%%%%%%%%%%%%%%%%%%%%%%%%%%%%%%%%%%%%%

\section{Discussion and Conclusions}

The electroweak data are well-fit by several of the patterns of 
neutrino-gauge coupling suppressions considered.  
In all cases considered, the best-fit values of the $\varepsilon_\ell$ are positive, 
\textit{i.e.} neutrino-gauge couplings are suppressed with respect to the Standard Model, 
as is required in models of neutrino mixing.
In models in which $\varepsilon_e$ is allowed to be non-zero (A-D,F) 
the fit to the data is good, and the fit improves if $\varepsilon_\mu$ 
is allowed to be non-zero as well (B,C).  
The fit quality is degraded for models in which $\varepsilon_e$ is set to zero (E,G,H), 
and the fit with $\varepsilon_\tau$ alone (H) is poor.  
In general the overall $\chi^2$ is insensitive to the presence of $\varepsilon_\tau$ 
as a degree of freedom in the fit.  
The data prefer the model with only $\varepsilon_e$ non-zero (F) 
to the model with only $\varepsilon_\mu$ non-zero (G).
 
Since the $\mu \rightarrow e \gamma$ data from MEGA \cite{mega}
demands that either $\varepsilon_e$ or $\varepsilon_\mu$
is strongly suppressed \cite{Shrock:1977,Glashow:2003wt,LORTW2}, 
$\varepsilon_e\neq 0$, $\varepsilon_\mu\approx 0$ seems to be
the solution preferred by current data.  
However, we stress that the inconsistency within the lepton universality data 
makes any such conclusion tentative.  
For example, fits using only the lepton universality constraint from
$\mu$ and $\tau$-decay, Eq.~(\ref{taudecay}), 
which is free of QCD uncertainties, indicate that the fit with only 
$\varepsilon_\mu$ is superior to a fit with only $\varepsilon_e$, as
shown in Table~\ref{TauFitResults}.
Therefore, future improvements in the lepton universality data 
(\textit{e.g.} better determination of the $\tau$ lifetime by Belle and Babar \cite{Gan:2003fj},
measurement of $B(\pi\rightarrow e\bar{\nu}_e)$ at the $0.2\%$ level by PIBETA \cite{pibeta}, \textit{etc.})
may ultimately provide a different conclusion.
In figure~\ref{fitCfig}(f),
the current 90\% contour overlaps with the $\varepsilon_e$ axis
but not with the $\varepsilon_\mu$ axis. If the region preferred
by the lepton universality data (dashed contour) is shifted toward the
$\varepsilon_\mu$ axis, $\varepsilon_e\approx0$, $\varepsilon_\mu\neq 0$ 
may become a viable solution also.

Langacker \cite{Langacker:2003tv} has noted that the observed 
violation of unitarity in the CKM matrix \cite{Abele:2002wc} will be aggravated
by suppressions of neutrino-gauge couplings. 
However, if the suppression parameters are allowed to break universality,
it is only a non-zero $\varepsilon_\mu$ that aggravates the CKM unitarity problem.
Thus the CKM unitary data actually prefers the $\varepsilon_e\neq 0$, $\varepsilon_\mu\approx 0$
solution (in the sense that it does not make the problem worse).
An improved determination of $|V_{ud}|$ is expected from the UCN-A experiment
at LANL in the near future \cite{Tipton:2000qi}.

The fits A, B, C, D, and F with excellent $\chi^2$'s require 
$T$ to be negative by more than $3\sigma$, 
$U$ to be positive by more than $3\sigma$, while
$S$ is within $1\sigma$ of zero.
As discussed in Ref.~\cite{LOTW1}, 
the $S$ and $T$ parameters can be accommodated within the Standard Model by 
increasing the Higgs mass to several hundred GeV.
The large $U$ parameter arises in part from discrepancy between the Standard Model 
prediction for the $W$ mass and in part from the shift due to the other fit parameters.
Neutrino mixing alone does not account for this discrepancy between the predicted and 
observed values of the $W$ mass; the $U$ parameter appears to require new physics.  
Whether a large $U$ parameter can be generated without a correspondingly
large $T$ parameter in some model is an open question that needs to be addressed.

The constraint on the suppression parameters
$\varepsilon_\ell$ ($\ell=e,\mu,\tau$) from
muon $g-2$ \cite{Sichtermann:2003cc,Ma:2002df} is weak \cite{rayyan}.
Further constraints may be obtained from 
$\mu$ to $e$ conversion in nuclei \cite{meco,Simkovic:2001fy}, and
muonium-antimuonium oscillation \cite{Willmann:1998gd,Aoki:aj,Halprin:wm}.
These will be discussed in a future publication.

%%%%%%%%%%%%%%%%%%%%%%%%%%%%%%%%%%%%%%%%%%%%%%%%%%%%%%%%%%%%%%%%%%%%%%%%%%%%%
%\newpage

\section*{Acknowledgments}

We would like to thank K. K. Gan, Gene Golowich, Randy Johnson, Alexey Pronin, and Kei Takeuchi for helpful communications and discussions.
This research was supported in part by the U.S. Department of Energy, 
grants DE--FG05--92ER40709, Task A (T.T., N.O., and S.R.),
and DE--FG02--84ER40153 (L.C.R.W.).

%%%%%%%%%%%%%%%%%%%%%%%%%%%%%%%%%%%%%%%%%%%%%%%%%%%%%%%%%%%%%%%%%%%%%%%%%%%%%
%\newpage

%%%%%%%%%%%%%%%%%%%%%%%%%%%%%%%%%%%%%%%%%%%%%%%%%%%
\newpage

\begin{table}[p]
\begin{tabular}{ll}
\begin{tabular}[t]{c|rrrr}
A & $S$ & $T$ & $U$ & $\varepsilon$ \\
\hline
$S$ & $\phantom{-}1.00$ & $0.56$ & $-0.21$ &  $0.18$ \\
$T$ &        & $\phantom{-}1.00$ & $-0.74$ & $-0.64$ \\
$U$ &        &        &  $\phantom{-}1.00$ &  $0.58$ \\
$\varepsilon$ &  &  &         &  $\phantom{-}1.00$ \\
\end{tabular}
\hfill &
\begin{tabular}[t]{c|rrrrrr}
B & $S$ & $T$ & $U$ & $\varepsilon_e$ & $\varepsilon_\mu$ & $\varepsilon_\tau$ \\
\hline
$S$ & $\phantom{-}1.00$ & $0.47$ & $-0.15$ &  $0.12$ &  $0.22$ & $-0.12$ \\
$T$ &        & $\phantom{-}1.00$ & $-0.77$ & $-0.60$ & $-0.34$ &  $0.08$ \\
$U$ &        &                   &  $1.00$ &  $0.54$ &  $0.33$ & $-0.09$ \\
$\varepsilon_e$ &  &  &                    &  $1.00$ & $-0.04$ & $-0.19$ \\
$\varepsilon_\mu$ & & &         &                    &  $1.00$ & $-0.09$ \\
$\varepsilon_\tau$ & & &        &                    &         &  $1.00$
\end{tabular}
\\
& \\
\begin{tabular}[t]{c|rrrrr}
C & $S$ & $T$ & $U$ & $\varepsilon_e$ & $\varepsilon_\mu$ \\
\hline
$S$ & $\phantom{-}1.00$ & $0.48$ & $-0.17$ &  $0.10$ &  $0.21$ \\
$T$ &        & $\phantom{-}1.00$ & $-0.77$ & $-0.59$ & $-0.33$ \\
$U$ &        &                   &  $1.00$ &  $0.53$ &  $0.33$ \\
$\varepsilon_e$ &  &             &         &  $1.00$ & $-0.06$ \\
$\varepsilon_\mu$ & &            &         &         &  $1.00$
\end{tabular}
\hspace{1cm} &
\begin{tabular}[t]{c|rrrrr}
D & $S$ & $T$ & $U$ & $\varepsilon_e$ & $\varepsilon_\tau$ \\
\hline
$S$ & $\phantom{-}1.00$ & $0.59$ & $-0.25$ &  $0.13$ & $-0.10$ \\
$T$ &        & $\phantom{-}1.00$ & $-0.74$ & $-0.65$ &  $0.05$ \\
$U$ &        &                   &  $1.00$ &  $0.58$ & $-0.06$ \\
$\varepsilon_e$ &  &             &         &  $1.00$ & $-0.19$ \\
$\varepsilon_\tau$ & &           &         &         &  $1.00$
\end{tabular}
\\
& \\
\begin{tabular}[t]{c|rrrrr}
E & $S$ & $T$ & $U$ & $\varepsilon_\mu$ & $\varepsilon_\tau$ \\
\hline
$S$ & $\phantom{-}1.00$ & $0.68$ & $-0.26$ &  $0.23$ & $-0.10$ \\
$T$ &        & $\phantom{-}1.00$ & $-0.67$ & $-0.45$ & $-0.04$ \\
$U$ &        &                   &  $1.00$ &  $0.42$ &  $0.02$ \\
$\varepsilon_\mu$ &  &           &         &  $1.00$ & $-0.10$ \\
$\varepsilon_\tau$ & &           &         &         &  $1.00$
\end{tabular}
\hspace{1cm} &
\begin{tabular}[t]{c|rrrr}
F & $S$ & $T$ & $U$ & $\varepsilon_e$ \\
\hline
$S$ & $\phantom{-}1.00$ & $0.60$ & $-0.25$ &  $0.12$ \\
$T$ &        & $\phantom{-}1.00$ & $-0.74$ & $-0.65$ \\
$U$ &        &        &  $1.00$ &  $0.59$ \\
$\varepsilon_e$ &  &  &         &  $1.00$ \\
\end{tabular}
\\
& \\
\begin{tabular}[t]{c|rrrr}
G & $S$ & $T$ & $U$ & $\varepsilon_\mu$ \\
\hline
$S$ & $\phantom{-}1.00$ & $0.68$ & $-0.26$ &  $0.22$ \\
$T$ &        & $\phantom{-}1.00$ & $-0.67$ & $-0.46$ \\
$U$ &        &        &  $1.00$ &  $0.42$ \\
$\varepsilon_\mu$ &  &  &         &  $1.00$ \\
\end{tabular}
\hfill &
\begin{tabular}[t]{c|rrrr}
H & $S$ & $T$ & $U$ & $\varepsilon_\tau$ \\
\hline
$S$ & $\phantom{-}1.00$ & $0.90$ & $-0.40$ & $-0.07$ \\
$T$ &        & $\phantom{-}1.00$ & $-0.59$ & $-0.10$ \\
$U$ &        &        &  $1.00$ &  $0.07$ \\
$\varepsilon_\tau$ &  &  &         &  $1.00$ \\
\end{tabular}
\\
\end{tabular}
\caption{Correlations among the fit parameters for fits A thought H.}
\label{epsCorr}
\end{table}

%%%%%%%%%%%%%%%%%%%%%%%%%%%%%%%%%%%%%%%%%%%%%%%%%%%
\newpage

%%%%%%%%%%%%%%%%%%%%%%%%%%%%%%%%%%%%%%%%%%%%%%%%%%%%%%%%%%%%%%%%%%
\newpage

%%%%%%%%%%%%%%%%%%%%%%%%%%%%%%%%%%%%%%%%%%%%%%%%%%%%%%%%%%%%%%%%%%
\end{document}